\documentclass[structabstract]{aa}  
\usepackage{graphicx}
\usepackage{float}
\usepackage{txfonts}
\usepackage{natbib}
\usepackage{longtable}
\usepackage[colorlinks=true,citecolor=cyan]{hyperref}
\def\pdeg{\ifmmode $\setbox0=\hbox{$^{\circ}$}\rlap{\hskip.11\wd0 .}$^{\circ}
  \else \setbox0=\hbox{$^{\circ}$}\rlap{\hskip.11\wd0 .}$^{\circ}$\fi}

\begin{document}
\title{The converging gas flow around the infrared dark cloud G28.37}


   \author{H.~Beuther
          \inst{1}
          \and
          C.~Gieser
          \inst{1}
          \and
          H.~Linz
          \inst{1}
          \and
          Q.~Zhang
          \inst{2}
          \and
          S.~Feng
          \inst{3}
          \and
          A.~Ahmadi
          \inst{4}
          \and
          J.D.~Soler
          \inst{5}
          \and
          D.~Semenov
          \inst{6,1}
          \and
          M.R.A.~Wells
          \inst{1}
          \and
          S.~Reyes-Reyes
          \inst{1}
}
   \institute{$^1$ Max Planck Institute for Astronomy, K\"onigstuhl 17,
     69117 Heidelberg, Germany, \email{name@mpia.de}\\
     $^2$ Harvard-Smithsonian Center for Astrophysics, 160 Garden St, Cambridge, MA 02420, USA\\
     $^3$ Department of Astronomy, Xiamen University, Zengcuoan West Road, Xiamen, 361005, People's Republic of China\\
     $^4$ ASTRON, Netherlands Institute for Radio Astronomy, Dwingeloo, the Netherlands\\
     $^5$ Department of Astrophysics. University of Vienna, T\"urkenschanzstrasse 17, 1180 Vienna, Austria\\
     $^6$  Zentrum für Astronomie der Universit\"at Heidelberg, Institut für Theoretische Astrophysik, Albert-Ueberle-Str. 2, 69120 Heidelberg, Germany
     }

   \date{Version of \today}

\abstract
    {How dense clouds and star-forming regions form out of the
      dynamical interstellar medium is at the heart of star formation
      research.}
    {The G28.37+0.07 star-forming region is a prototypical infrared dark cloud
      (IRDC) located at the interface of a
      converging gas flow. This study characterizes the properties of
      this dynamic gas flow.}
    {Combining data from the Northern Extended Millimeter Array
      (NOEMA) with single-dish data from the IRAM\,30\,m observatory, we mapped large spatial scales ($\sim$81\,pc$^2$) at high
      angular resolution ($7.0''\times 2.6''$ corresponding
      $\sim$2.3$\times 10^4$\,au or $\sim$0.1\,pc) down to core
      scales. The spectral setup in the 3\,mm band covers many
      spectral lines as well as the continuum emission.}
{The data clearly reveal the proposed west-east converging gas flow in
  all observed dense gas tracers. We estimate a mass-flow rate along
  that flow around $10^{-3}$\,M$_{\odot}$yr$^{-1}$. Comparing
  these west-east flow rates to infall rates toward sources along the
  line of sight, the gas flow rates are roughly a factor
  of 25 greater than than those along the line of sight (roughly
  perpendicular to the west-east flow). This confirms the dominance of
  longitudinal motions along the converging gas flow in this
  region. For comparison, in the main north-south IRDC formed
  by the west-east converging gas flow, infall rates along the
  line of sight are about an order of magnitude greater than those along the
  west-east flow. In addition to the kinematic analysis, a comparison of
  CH$_3$CN-derived gas temperatures with Herschel-derived dust
  temperatures typically show higher gas temperatures toward
  high-density sources. We discuss whether mechanical heating from the
  conversion of the flow's kinetic energy into thermal energy may
  explain some of the observed temperature differences.}
{Our analysis of the G28.37+0.07 converging gas flow shows that such
  structures can indeed form and feed typical high-mass
  star-forming regions in the Milky Way. The differences between flow
  rates along the converging flow, perpendicular to it, and toward
  the sources at the IRDC center indicate that at the
  interfaces of converging gas flows -- where most of the active star
  formation takes place -- originally more directed gas flows can
  convert into multidirectional infall motions.
}


\titlerunning{Dynamics and fragmentation in the infrared dark cloud G28.37+0.07}

\maketitle
 
\section{Introduction}
\label{intro}

The current paradigm for molecular cloud formation and the subsequent
collapse of star-forming regions into individual stars and clusters
views the entire process as a hierarchical gas flow from large to small
scales (e.g.,
\citealt{vazquez2000,vazquez2011,vazquez2017,vazquez2019,heitsch2005,glover2007,banerjee2009,ibanez2017,motte2018,ballesteros2018,padoan2020,pineda2023,hacar2023,beuther2025}). While
this picture is theoretically appealing, observational signatures are
still scarce, with notable exceptions presented in
\citet{schneider2010}, \citet{csengeri2011a,csengeri2011b},
\citet{hennemann2012}, \citet{tackenberg2014}, \citet{hacar2017},
\citet{beuther2020}, \citet{kido2025}, and
\citet{gieser2025}. Certainly, observational
studies following the gas flow from the largest cloud scales --
potentially still atomic -- to the smallest core scales, where individual
or multiple stars form, barely exist. These kinematic flows are directly associated
with the fragmentation of clouds as well as the chemical evolution
of the interstellar medium (ISM) and star-forming cores. A
central goal in molecular cloud and star formation research is to
characterize the entire gas flow process, the fragmentation, and the chemical evolution of the material in depth. This requires
covering a broad range of spatial scales as well as gas densities.

To address these topics, we embarked on a concerted effort to
investigate the prototypical infrared dark cloud (IRDC) G28.37+0.07 (G28 in
the following) from the atomic gas at large scales ($>$20\,pc,
\citealt{beuther2020}) via dense gas studies at typical single-dish
resolutions \citep{tackenberg2014} down to scales of several thousand au
only reachable with interferometers such as NOEMA.

\begin{figure*}[htb]
\includegraphics[width=0.5\textwidth]{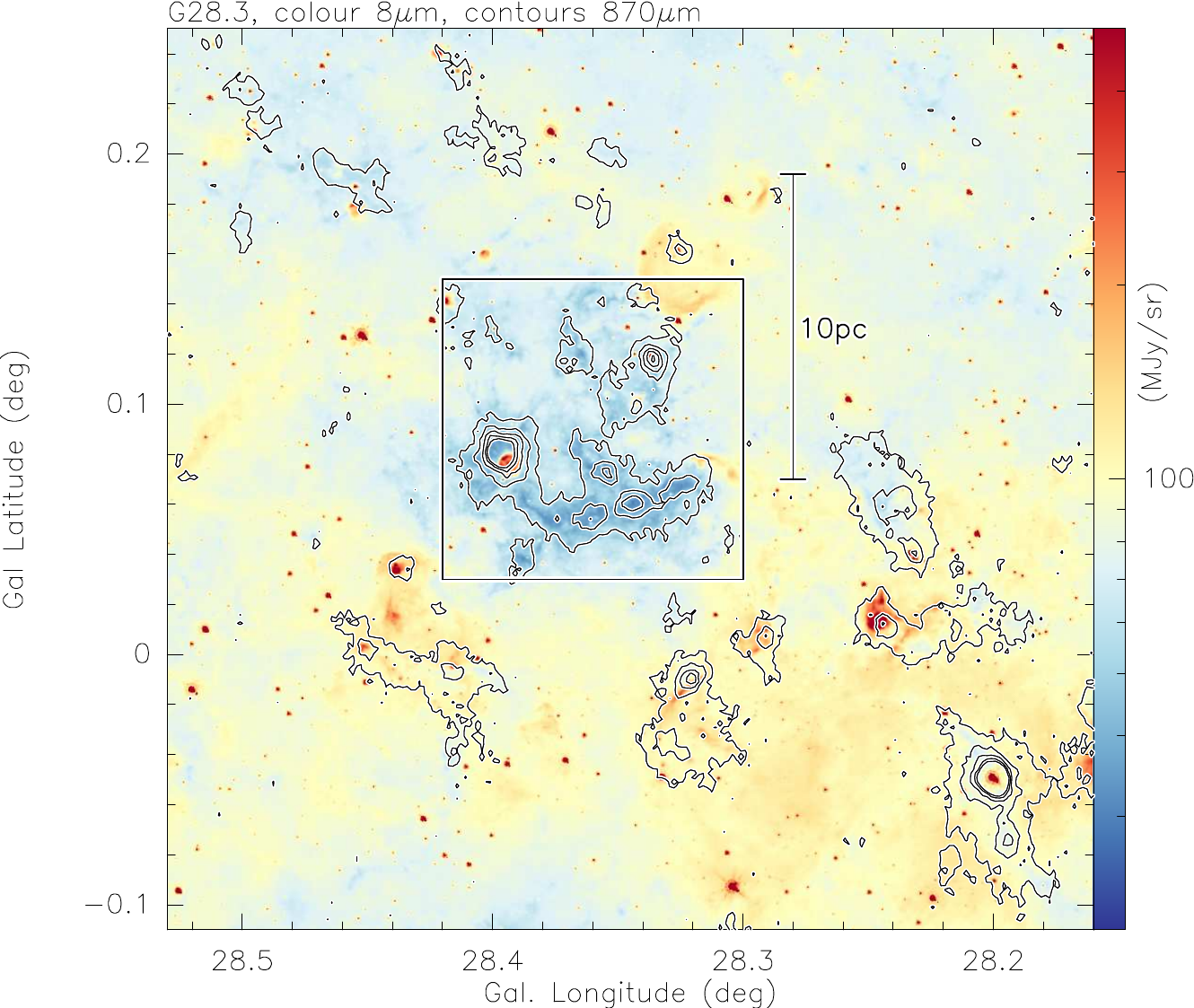}
\includegraphics[width=0.48\textwidth]{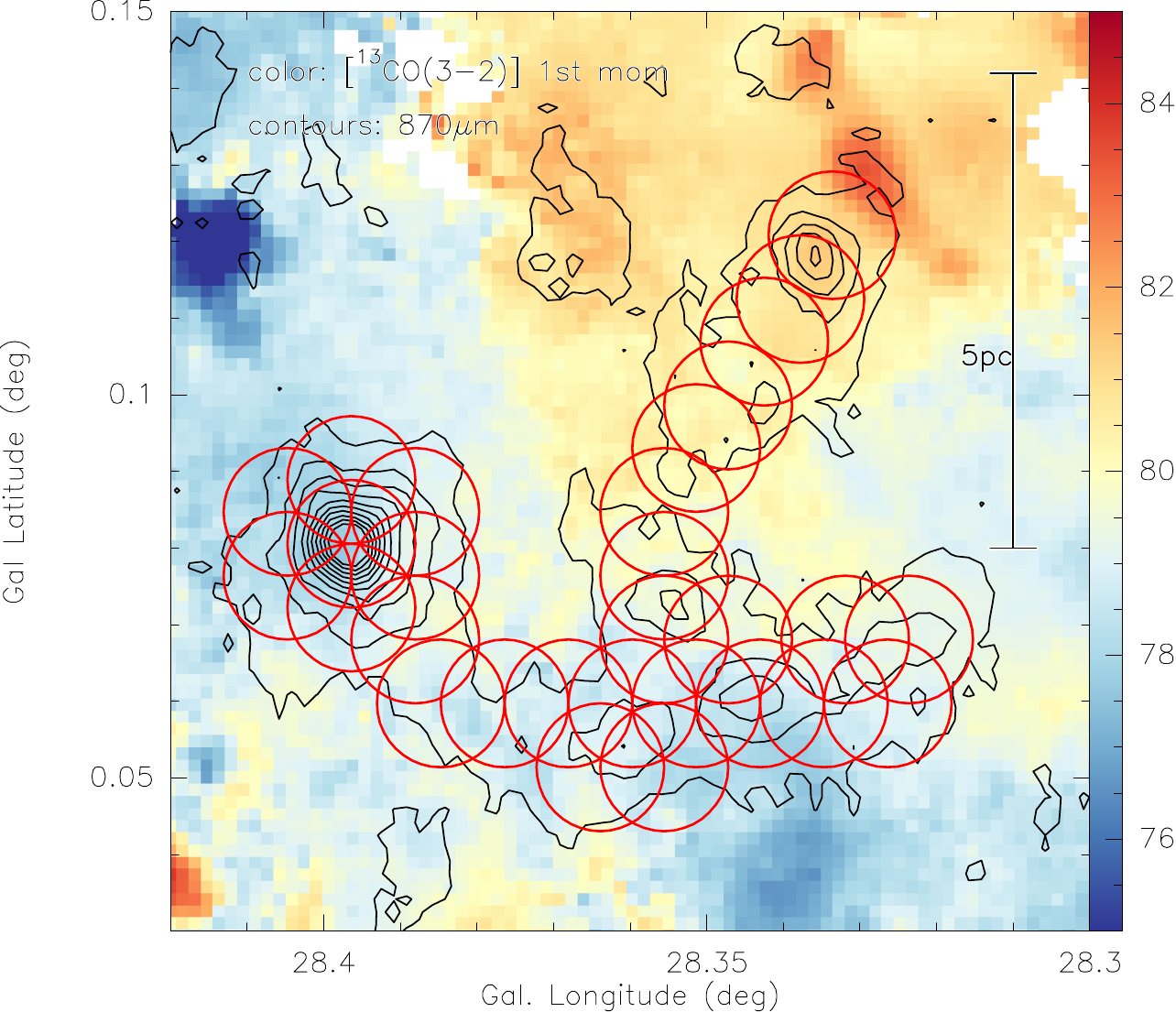}
\caption{Overview of the G28 region. Left panel: Color scale showing GLIMPSE 8\,$\mu$m emission \citep{churchwell2009}
  for the G28 IRDC. The contours show the corresponding ATLASGAL
  870\,$\mu$m dust continuum emission, starting at 4$\sigma$ (200\,mJy\,beam$^{-1}$) \citep{schuller2009} and continuing in
  8$\sigma$ steps. The box outlines the area shown to the right. 
    Right panel: Zoom into the G28 target region. The color scale
  presents the velocity field (first moment) observed in $^{13}$CO(3–2)
  with APEX \citep{beuther2020}. The contours show the
  870\,$\mu$m continuum emission from the ATLASGAL survey
  \citep{schuller2009}. The red circles outline the observed NOEMA
  mosaic. A scale bar is shown to the right in both panels.}
\label{overview} 
\end{figure*} 

The star formation region G28 at a distance of 4.7\,kpc is a large
cloud complex where the innermost region is the well-known IRDC G28
(e.g., \citealt{pillai2006,rathborne2006,wang2008,butler2012,tan2013,tackenberg2014,kainulainen2013b,zhang2015,feng2016b,kong2019,beuther2020,barnes2023,law2024,liu2024}). Infrared
data from the Spitzer satellite (Fig.~\ref{overview} left panel;
\citealt{churchwell2009}) show that filamentary extinction structures
extend from the large-scale outskirts down to the cloud's central
region. Observations of the dense gas in N$_2$H$^+$ indicate that the
region is in a stage of global collapse
\citep{tackenberg2014}. Different evolutionary stages of star
formation activity exist throughout the cloud from almost quiescent
clumps to bright mid-infrared high-mass protostellar objects (e.g.,
 \citealt{pillai2006,wang2008,butler2012,feng2016b}). This makes the
G28 complex an ideal candidate for investigating the kinematic and
chemical properties during cloud and star formation in a uniform way.

To investigate the kinematics on large scales ($>$30\,pc) from the
atomic and molecular envelopes of the IRDC, \citet{beuther2020} made
use of the HI self-absorption (HISA) data from the THOR survey
\citep{beuther2016}, $^{13}$CO(1--0) molecular line data from the
Galactic Ring survey \citep{jackson2006}, as well as APEX observations
in the atomic [CI] and $^{13}$CO(3--2) emission lines. This combined
dataset shows velocity gradients across the entire cloud complex
(Fig.~\ref{overview} right panel) that are interpreted as converging
gas flows with flow rates between 10$^{-5}$ and
10$^{-4}$\,M$_{\odot}$yr$^{-1}$ \citep{beuther2020}.
  
How do these large-scale characteristics relate to the small-scale
kinematics, fragmentation, and chemical evolution of the central G28
IRDC? We address these topics here via a NOEMA (Northern
Extended Millimeter Array) mosaic study of the central G28 IRDC.

\section{Observations}
\label{obs}

\subsection{The Northern Extended Millimeter Array (NOEMA)}

The G28 IRDC was observed between June and October 2019 with NOEMA in
the 3\,mm band with nine antennas in the array. The two compact
configurations, D and C, were used, and we observed a 29-field mosaic,
as outlined in Figure \ref{overview} (right panel). Phase calibration
was conducted through regularly interleaved observations of quasars
1829-106 and 1851+0035. For bandpass calibration, we used the strong
quasar 1749+096, and for the flux calibration we observed MWC349. The
calibration was conducted in {\sc clic}, and imaging was subsequently
performed in {\sc mapping}; both programs are part of the GILDAS
software suite\footnote{http://www.iram.fr/IRAMFR/GILDAS}. The phase
center of the mosaic is Right Ascension~(J2000.0) 18:42:45.65,
Declination~(J2000.0) $-$04:01:47.75. The rest velocity of the region
is $v_{\rm LSR}=79.5$\,km\,s$^{-1}$ (e.g., \citealt{beuther2020}). The
3\,mm tuning covered a broad bandpass between 70.22 to 78.33\,GHz in
the lower sideband and between 85.71 to 93.82\,GHz in the upper
sideband. While the spectral resolution over the entire bandpass is
2\,MHz, the polyfix correlator is extremely flexible by accommodating
many high-spectral resolution sub-windows at a resolution of
62.5\,kHz. For the analysis conducted in this paper, we restricted
ourselves to a suite of lines tracing different densities, deuterated
and non-deuterated species, as well as shock tracers. The spectral
lines and parameters are listed in Table \ref{lines}.

\begin{figure*}[htb]
\includegraphics[width=0.99\textwidth]{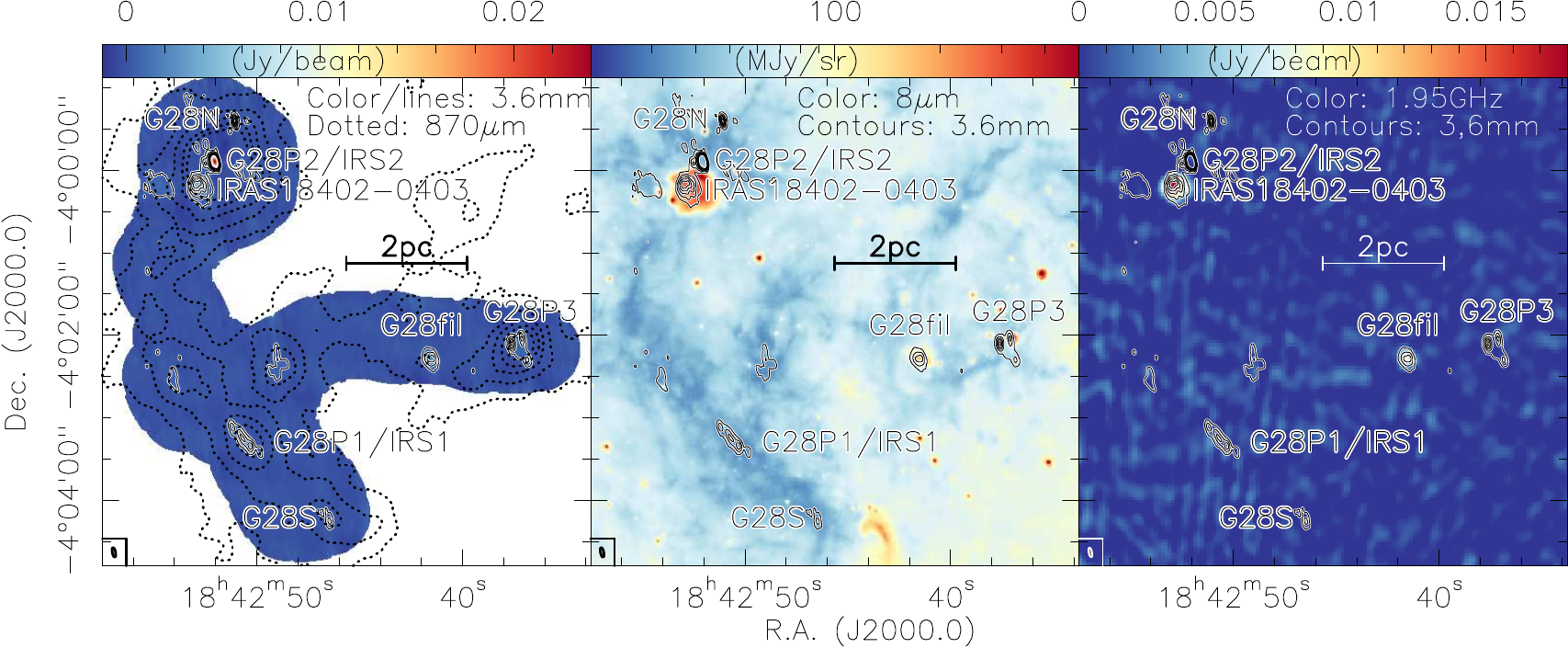}
\caption{Continuum images of the G28 IRDC. Color scale
  and solid contours showing the NOEMA 3\,mm continuum emission. Contour
  levels range from 0.4 to 1.6\,mJy\,beam$^{-1}$ ($1\sigma \sim
  0.1$\,mJy\,beam$^{-1}$).  The dotted contours show 870\,$\mu$m
  single-dish continuum data (ATLASGAL, \citealt{schuller2009})
  starting at 0.15\,Jy\,beam$^{-1}$ and continuing in steps of   0.3\,Jy\,beam$^{-1}$ up to 1.5\,Jy\,beam$^{-1}$ ($1\sigma \sim
  0.05$\,Jy\,beam$^{-1}$).  Middle: Color scale showing the
  Spitzer 8\,$\mu$m emission ($2''$ resolution,
  \citealt{churchwell2009}) with 3\,mm emission overlaid as contours. 
    Right: Color scale showing the 1.95\,GHz continuum emission
  from the THOR survey ($10.6''\times 9.0''$ resolution,
  \citealt{beuther2016,wang2020a}) with 3\,mm emission overlaid as
  contours. All panels show a scale bar and the NOEMA synthesized
  beam. Source labels follow \citet{carey2000}, \citet{wang2008}; G28N
  and G28fil are newly labeled here.}
\label{continuum} 
\end{figure*} 

While the spectral line data were imaged after complementing the short
spacings from the IRAM 30\,m observatory (see the following
subsections), continuum short spacing observations for this region do
not exist, and we imaged NOEMA-only data. After excluding the 29
strongest spectral lines, the entire lower and upper sideband (LSB and
USB) data were collapsed into LSB and USB continuum data and then
merged into double-sideband (DSB) continuum data used in the
following analysis. The final DSB continuum was created from $\sim$15.69\,GHz
of the bandpass. Using natural weighting and hogbom cleaning, the
resulting beam size and rms are $7.0''\times 2.6''$ and
0.1\,mJy\,beam$^{-1}$.

\begin{table}[htb]
\caption{Line parameters}
\begin{tabular}{lrrrrr}
  \hline \hline
line & freq. & $1\sigma_{\rm 30m}$ & $1\sigma_{\rm merged}$ & beam & $n_{\rm eff}^a$ \\
& (GHz) & $\left(\frac{\rm Jy}{\rm beam}\right)$ & $\left(\frac{\rm mJy}{\rm beam}\right)$ & $('')$ & $\left(\frac{10^3}{\rm cm^3}\right)$ \\
\hline
continuum                 & 82.02 &   --   & 0.1 & $7.0\times 2.6$ & \\
DCO$^+$(1--0)             & 72.039 & 0.187 & 14.5 & $8.3\times 3.2$ & \\
SO$_2(6_{0,6}-5_{1,5})$    & 72.758 & 0.155 & 13.0 & $8.2\times 3.1$ & \\
H$_2$CO$(1_{0,1}-0_{0,0})$ & 72.838 & 0.161 & 9.9 & $8.2\times 3.1$ & 2.6 \\
CH$_3$CN$(4_0-3_0)$       & 73.590 & 0.144 & 9.5 & $8.2\times 3.1$ & 37\\
DNC(1--0)                & 76.306  & 0.111 & 9.9 & $8.0\times 3.0$ & \\
NH$_2$D(1--0)            & 85.926  & --    & 7.5 & $7.0\times 2.5$ & \\
H$^{13}$CO$^+$(1--0)      & 86.754 & 0.165 & 7.6 & $7.3\times 2.6$ & 22\\
SiO(2--1)                & 86.847 & 0.149 & 8.6 & $7.3\times 2.6$ & 15$^b$ \\
HCN(1--0)                & 88.632 & 0.146 & 16.1 & $7.2\times 2.6$& 4.5 \\
HCO$^+$(1--0)            & 89.189 & 0.160 & 9.5 & $7.1\times 2.5$ & 0.5 \\
CH$_3$CN$(5_0-4_0)$      & 91.987 & 0.148 & 6.8 & $7.0\times 2.5$ & 74 \\
$^{13}$CS(2--1)          & 92.494 & 0.126 & 9.6 & $6.9\times 2.5$ & 5.4$^c$\\
N$_2$H$^+$(1--0)         & 93.174 & 0.146 & 11.8 & $6.9\times 2.5$ & 5.5\\
\hline \hline
\end{tabular}
~\\ Notes: The $1\sigma$ rms values are for all lines in the merged
data at 0.8\,km\,s$^{-1}$ channel width. Only NH$_2$D is for the
NOEMA-only data. The continuum rms is for NOEMA-only data across the whole
bandpass, excluding the strong spectral lines.\\ $^a$ Effective
densities at 20\,K from \citet{shirley2015}.\\ $^b$ Critical densities
(no effective densities in \citet{shirley2015}).\\ $^c$ Value for
CS(1--0) in \citet{shirley2015}.
\label{lines}
\end{table}

\subsection{IRAM 30\,m}

To complement the missing short spacings of the spectral line data and
obtain large-scale information, we observed the same region in the
on-the-fly mode with the IRAM\,30\,m telescope within project 100-19
between February and July 2020. The region was scanned in Right
Ascension and Declination to minimize any artificial striping
effects. The beam of the 30\,m at 86\,GHz is roughly $29''$. The
FTS200 correlator with 0.2\,MHz spectral resolution (corresponding to
0.8\,km\,s$^{-1}$ at 75\,GHz) was employed. For the strongest lines,
we used the even higher spectral-resolution unit with 50\,kHz
resolution. In the end, to achieve a homogeneous spectral resolution
for all lines, we used 0.8\,km\,s$^{-1}$ for the single-dish as well
as the merged NOEMA+30\,m datasets. Following IRAM guidelines, calibration to main beam temperature $T_{\rm mb}$ was performed by multiplying
the data by the ratio of forward efficiency to beam efficiency $f_{\rm
  eff}/b_{\rm eff}=1.17${\footnote{http://www.iram.es/IRAMES/mainWiki/Iram30mEfficiencies}. The
conversion to Jy\,beam$^{-1}$ was estimated in the Rayleigh-Jeans
limit at the respective frequencies taking into account the beam
size. This conversion factor was typically around 4.94. The $1\sigma$
rms values are also reported in Table \ref{lines}.

\subsection{Merged data}

The merging of the NOEMA and 30\,m data was again conducted in {\sc
  mapping}, and the 30\,m data were converted into pseudo-visibilities
via the {\sc uv\_short} task. The NOEMA and 30\,m data were then
imaged together with natural weighting, hogbom cleaning, and
0.8\,km\,s$^{-1}$ channel width. The final rms of the merged data
$1\sigma_{\rm merged}$ and the resulting beam sizes are reported in
Table \ref{lines}.

\section{Results}
\label{results}

\subsection{Continuum emission}
\label{continuum_section}

Fig.~\ref{continuum} presents the new NOEMA 3\,mm continuum data
compared to the single-dish 870\,$\mu$m data \citep{schuller2009},
mid-infrared Spitzer 8\,$\mu$m data \citep{churchwell2009}, and centimeter
wavelength 1.95\,GHz emission from the THOR survey
\citep{beuther2016,wang2020a}. The mid-infrared 8\,$\mu$m data clearly
show that the long north-south filamentary structure is mainly
infrared-dark, indicating the overall youth and high density of the
region. Nevertheless, to the north lies the bright infrared and centimeter
continuum source IRAS18402-0403, which is clearly in a more advanced
evolutionary stage. Directly north of that lies the
G28P2/IRS2 source, which is barely detected in the 8\,$\mu$m Spitzer
image (but nevertheless hosts an infrared source also detected with
JWST; Reyes-Reyes et al.~subm.~to A\&A) and not detected at centimeter
wavelengths. Regarding the east-west structure, while partly also
infrared-dark, G28fil has infrared and centimeter
counterparts toward the west, and G28P3 has an infrared counterpart as well. Hence, while overall young and often classified as a prototypical infrared-dark cloud, the region clearly hosts a diverse range of evolutionary
stages. Even the infrared-dark structures are largely not
starless; prominent detection of many outflows confirms the
presence of active star formation in these regions as well (e.g.,
\citealt{wang2006,wang2012c,feng2016b,tan2016,kong2019}).

\begin{figure}[htb]
\includegraphics[width=0.49\textwidth]{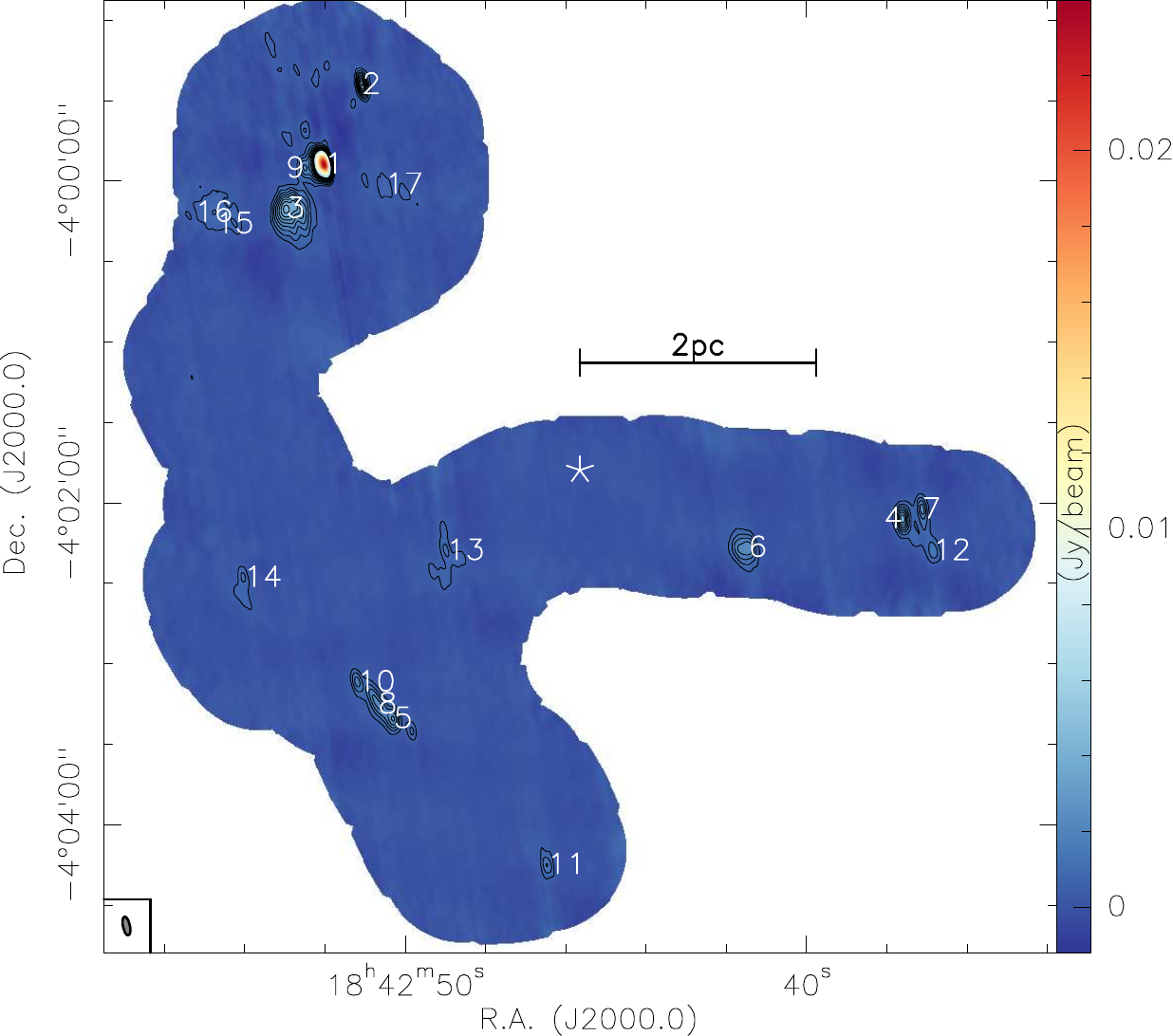}
\caption{3.6\,mm continuum data with source identifications. The
  contour levels start at the 5$\sigma$ level (0.5\,mJy\,beam$^{-1}$) and continue in 5$\sigma$ steps. The sources from the
  Clumpfind identification are labeled; the star marks the
  phase-center position for the source offsets in Table
  \ref{cont}. A scale bar and the synthesized beam are shown as well.}
\label{finding} 
\end{figure} 

\begin{table*}[htb]
  \caption{Continuum parameters}
  \label{cont}
  \begin{tabular}{lrrrrrrrrrrrr}
    \hline \hline
    \# & R.A. & Dec. & $S_{\rm{peak}}$ & $S_{\rm{int}}$ & $R$ & $T_{\rm dust}$ & $T_{\rm CH_3CN}$ & $T_{\rm CH_3CN}$ & $M_{20K}$ & $N_{20K}$ & $M_{T_{\rm CH_3CN}}$ & $N_{T_{\rm CH_3CN}}$ \\
    & (J2000) & (J2000) & $\left(\frac{\rm mJy}{\rm beam}\right)$ & (mJy) & (pc) & (K) & (K) & (@$12''$, K) & ($M_{\odot}$) & ($10^{23}$cm$^{-2}$) & ($M_{\odot}$) & ($10^{23}$cm$^{-2}$) \\
    \hline
1  &  18:42:52.02 & -03:59:54.0 &  23.97 &  39.8 &  0.14 & 17 & 41 & 33 & 1692 & 42.9  & 784 &  19.9 \\
2  &  18:42:51.07 & -03:59:24.1 &   5.09 &   5.6 &  0.10 & 17 & -- & -- & 237  &  9.1  &  -- &   --  \\
3$^a$&18:42:52.97 & -04:00:10.3 &   4.22 &  26.7 &  0.22 & 19 & -- & -- &  --  &  --   &  -- &   --  \\
4  &  18:42:37.61 & -04:02:06.0 &   3.87 &   7.2 &  0.13 & 22 & 30 & 21 & 309  &  6.9  & 199 &   4.5 \\
5  &  18:42:50.28 & -04:03:20.6 &   2.62 &   4.2 &  0.10 & 17 & 24 & 17 & 179  &  4.7  & 146 &   3.8 \\
6$^a$&18:42:41.46 & -04:02:16.8 &   2.46 &   8.4 &  0.15 & 20 & -- & -- & --   &  --   &  -- &   --  \\
7  &  18:42:37.12 & -04:02:02.3 &   2.31 &   3.8 &  0.11 & 23 & 21 & 19 & 159  &  4.1  & 151 &   3.9 \\
8  &  18:42:50.67 & -04:03:15.1 &   2.18 &   5.1 &  0.12 & 17 & 22 & 20 & 218  &  3.9  & 197 &   3.5 \\
9  &  18:42:52.49 & -03:59:55.2 &   2.13 &   4.1 &  0.11 & 17 & -- & -- & 175  &  3.8  &  -- &   -- \\
10 &  18:42:51.19 & -04:03:07.0 &   1.73 &   2.5 &  0.09 & 17 & -- & -- & 108  &  3.1  &  -- &   -- \\
11 &  18:42:46.45 & -04:04:15.0 &   1.53 &   1.9 &  0.08 & 17 & -- & -- & 79   &  2.7  &  -- &   -- \\
12 &  18:42:36.86 & -04:02:17.8 &   1.43 &   2.9 &  0.11 & 19 & -- & -- & 122  &  2.6  &  -- &   -- \\
13 &  18:42:48.97 & -04:02:17.8 &   1.24 &   3.9 &  0.14 & 17 & 26 & 22 & 166  &  2.2  & 125 &   1.7 \\
14 &  18:42:54.02 & -04:02:28.0 &   1.16 &   2.2 &  0.10 & 18 & -- & -- & 92   &  2.1  &  -- &   -- \\
15 &  18:42:54.24 & -04:00:16.0 &   1.08 &   2.0 &  0.10 & 18 & -- & -- & 83   &  1.9  &  -- &   -- \\
16 &  18:42:54.77 & -04:00:11.8 &   1.02 &   4.5 &  0.15 & 18 & -- & -- & 190  &  1.8  &  -- &   -- \\
17 &  18:42:50.51 & -04:00:01.5 &   0.83 &   1.1 &  0.08 & 17 & -- & -- & 47   &  1.5  &  -- &   --\\
    \hline \hline
  \end{tabular}
~\\ Notes: $^a$ Significant free-free contribution (Fig.~\ref{continuum}), with no mass and column density estimates.
\end{table*}

Using the Clumpfind algorithm \citep{williams1994} on the 3.6\,mm
continuum data, we extracted sources from the region above a $5\sigma$
threshold. Peak and integrated fluxes are presented in Table
\ref{cont}. The source numbers are sorted in decreasing peak
intensities $S_{\rm peak}$, and Fig.~\ref{finding} presents the
respective source number assignments. Assuming optically thin dust
continuum emission at a 3.6\,mm wavelength, we can estimate the column
densities $N$ and masses $M$ following \citet{hildebrand1983} and
\citet{schuller2009}. We assumed a gas-to-dust-mass ratio of 150
\citep{draine2011} and a dust opacity $\kappa=0.1$\,cm$^2$g$^{-1}$,
extrapolated from \citet{ossenkopf1994} to 3.6\,mm at densities of
$10^5$\,cm$^{-3}$. We did not derive column density and mass estimates
for IRAS18402-0403 (source \#3) and G28fil (source \#6) since
free-free emission (Fig.~\ref{continuum}, right panel) strongly
contaminates the 3.6\,mm fluxes .

Regarding the temperatures of the region, we can follow two
approaches. On the one hand, the dust temperature map derived by
\citet{marsh2017} from the Herschel far-infrared data reveals
temperatures at an angular resolution of $12''$ toward all sources
between 17 and 23\,K (Table \ref{cont} \& Fig.~\ref{temp}). On the
other hand, our spectral setup covers two CH$_3$CN $k$-ladders: CH$_3$CN$(5_k-4_k)$ and CH$_3$CN$(4_k-3_k)$ with $k$ from
0 to 4. Assuming local thermodynamic equilibrium, both $k$-ladders can
be fit together to derive a gas temperature map of the region using
the XCLASS radiative transfer package \citep{moeller2017}. We assume
that the lines originate from one emission component. The noise was
evaluated at each pixel and only pixels with signal-to-noise ratio
$\geq 6$ were fit (e.g.,
\citealt{gieser2021,gieser2022}). Furthermore, high-noise regions
corresponding to the noisy areas at the edge of the field of view
($\geq 0.2$\,K) were masked as well. Figure \ref{temp} presents a
comparison of the derived temperature maps. While the dust
temperatures can be estimated for the entire field, CH$_3$CN is only
detected toward the brightest continuum sources and hence is less
extended. Nevertheless, where detected, the CH$_3$CN-derived gas
temperatures are consistently higher than those derived from the
far-infrared continuum emission. After smoothing the CH$_3$CN data to
the $12''$ resolution of the Herschel temperature map and running
XCLASS on these data again (Fig.~\ref{temp} right panel), some
temperature differences became smaller; nevertheless, for G28P2/IRS2 (source 1)
the estimated temperatures still varied by almost a factor of 2. Table
\ref{cont} lists the source-averaged mean temperatures for the detected
sources from the dust, CH$_3$CN, and smoothed CH$_3$CN data.  We revisit this potential decoupling of gas and dust temperatures in
subregions of G28 in Sect.~\ref{decouple}.

\begin{figure*}[ht!]
\includegraphics[width=0.99\textwidth]{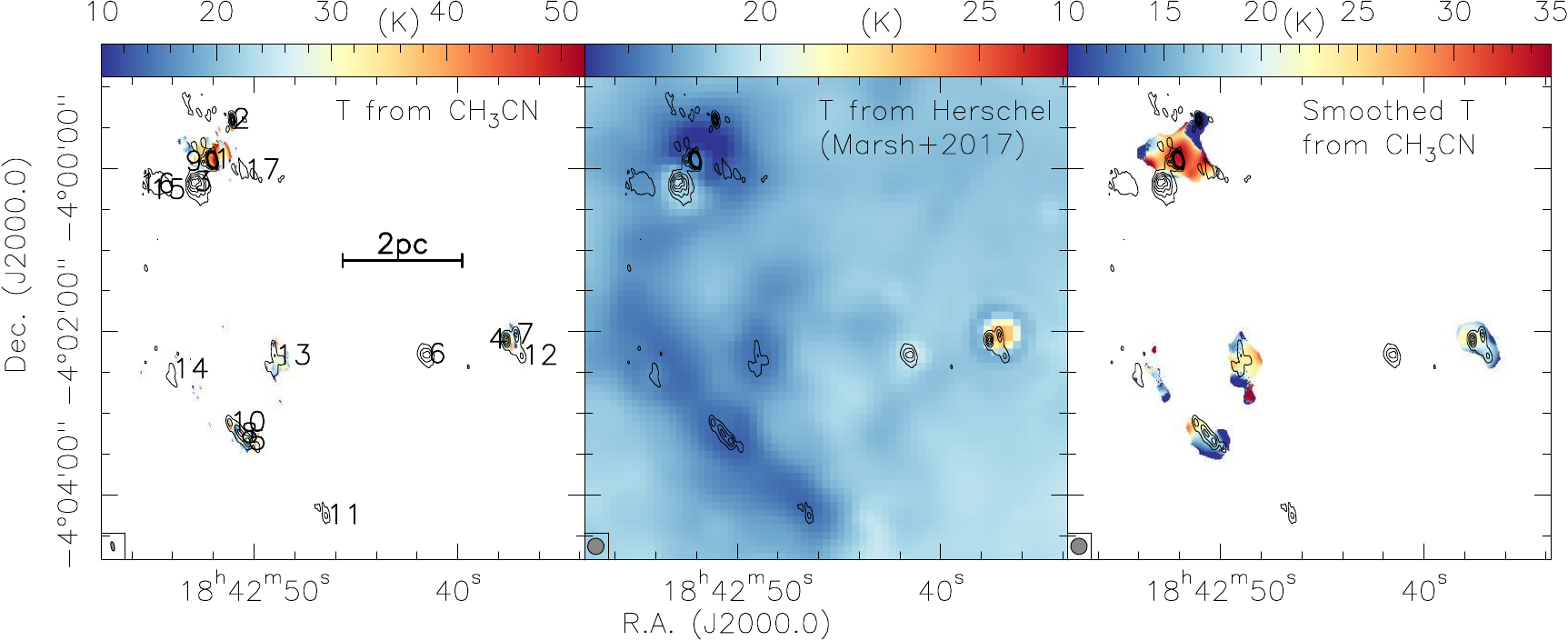}
\caption{Temperature maps. Left: Temperature map from combined fitting of CH$_3$CN$(5_k-4_k)$ and
  $(4_k-3_k)$ $k$-ladders ($k$ from 0 to 4). Middle: Dust temperature map from Herschel far-infrared
  data \citep{marsh2017}. Right: CH$_3$CN temperature map, smoothed to
  the same $12''$ spatial resolution of the Herschel map. The
  contours outline the NOEMA 3.6\,mm continuum emission; contour
  levels are from 0.4 to 1.6\,mJy\,beam$^{-1}$ ($1\sigma \sim
  0.1$\,mJy\,beam$^{-1}$). The source numbers and a scale bar are shown
  in the left panel.}
\label{temp} 
\end{figure*} 

For the mass and column densities, we used both temperatures. Since the
dust temperatures are all close to 20\,K, we estimated the masses and
column densities once for all sources using a single temperature of 20\,K. In addition, for sources with CH$_3$CN detections, we also
estimated these parameters using the CH$_3$CN-derived temperatures at
the original resolution, which corresponds to that of the continuum
data. All results are listed in Table \ref{cont}.  The source
masses derived this way from the 3.6\,mm continuum data for the G28
IRDC range broadly from 47 to more than 1600\,M$_{\odot}$
(or up to 784\,M$_{\odot}$ when using the CH$_3$CN
temperatures). With the single-dish ATLASGAL 870\,$\mu$m continuum
data shown in the left panel of Fig.~\ref{continuum}, we can also
estimate the total mass of the region. Using the same parameters as
above and assuming 20\,K once again, with a dust opacity
$\kappa=1.85$\,cm$^2$g$^{-1}$ at 870\,$\mu$m \citep{schuller2009}, the
total gas mass of the region is $\sim 1.26\times 10^4$\,M$_{\odot}$
(within a factor of 2 of the Taurus cloud mass,
\citealt{goldsmith2008}).  Comparing this to the total mass of all
identified sources from the NOEMA data, we recover roughly 31\% of the emission and mass using
the interferometer. The column
densities cover a similar range from around $10^{23}$\,cm$^{-2}$
to roughly $4\times 10^{24}$\,cm$^{-2}$ (or $\sim$2.0$\times
10^{24}$\,cm$^{-2}$ when using the CH$_3$CN temperatures). Estimated
effective source radii are larger than 0.08\,pc (or
$>$16500\,au), at the lower end roughly at our spatial resolution
limit. For comparison, higher angular resolution dust continuum images
for different subregions indicate even more fragmentation at
smaller scales (see, e.g., \citet{zhang2015} and \citet{kong2019}).

\subsection{Integrated molecular line emission}

In the following, we concentrate on the spectral line emission and the
associated kinematic properties of the gas. Fig.~\ref{mom0} shows a
compilation of integrated intensity maps for our selected molecular
lines at low ($\sim$29$''$) and high ($7''\times 2.5''$, Table
\ref{lines}) angular resolutions. Starting with the 30\,m data, all
the presented molecular lines are well detected in the entire area of
$\sim 9\times 9$\,pc$^2$. While this is not surprising for
abundant species such as HCO$^+$, H$_2$CO, and N$_2$H$^+$, as well as deuterated
species such as DCO$^+$ and DNC, shock tracers such as SiO and SO$_2$ or
higher density tracers such as CH$_3$CN and $^{13}$CS emit extended
emission toward several clumps in the region. The main strong
molecular line emission structures are G28P2/IRS2 and IRAS18402-0403
in the north, G28P1/IRS1 and G28S in the south, and G28P3 in the west
(see Fig.~\ref{continuum} for the labeling).

Zooming into smaller spatial scales with the merged NOEMA+30\,m data
(Fig.~\ref{mom0}, bottom panels), the general structures are similar as
expected. However, the higher angular resolution clearly
reveals that most of the molecular emission in the north does not stem
from the H{\sc ii} region IRAS18402-0403 but is emitted from the
younger region G28P2/IRS2. The only exception in that regard is the
HCN map that also shows emission around the H{\sc ii} region. Since it
is more diffuse and not very peaked, that emission likely stems from
the larger-scale clump environment of IRAS18402-0403.

Regarding deuterated species, although the 30\,m-only data show extended
DCO$^+$ emission, the NOEMA+30\,m map is comparably noisy without
strong compact emission structures. Inspecting the NOEMA-only
DCO$^+$ data reveals barely any detectable features. This indicates
that DCO$^+$ in this region is found more in the diffuse environmental
gas at lower temperatures and less toward the peak-structures. For
comparison, NH$_2$D -- which was not observed with the 30\,m leaving
only NOEMA data -- shows compact emission toward the two youngest
sources G28P1/IRS1 and G28S, whereas in the north
emission structures appear at the edges of G28P2/IRS2 and IRAS18402-0403. The
deuterated DNC map lies between the other two; it shows substantial extended
emission but also some compact structures (see also \citet{feng2019}
for a deuteration study of G28P1 and G28S).

\begin{figure*}[htb]
\hspace{0.4cm} \includegraphics[width=0.87\textwidth]{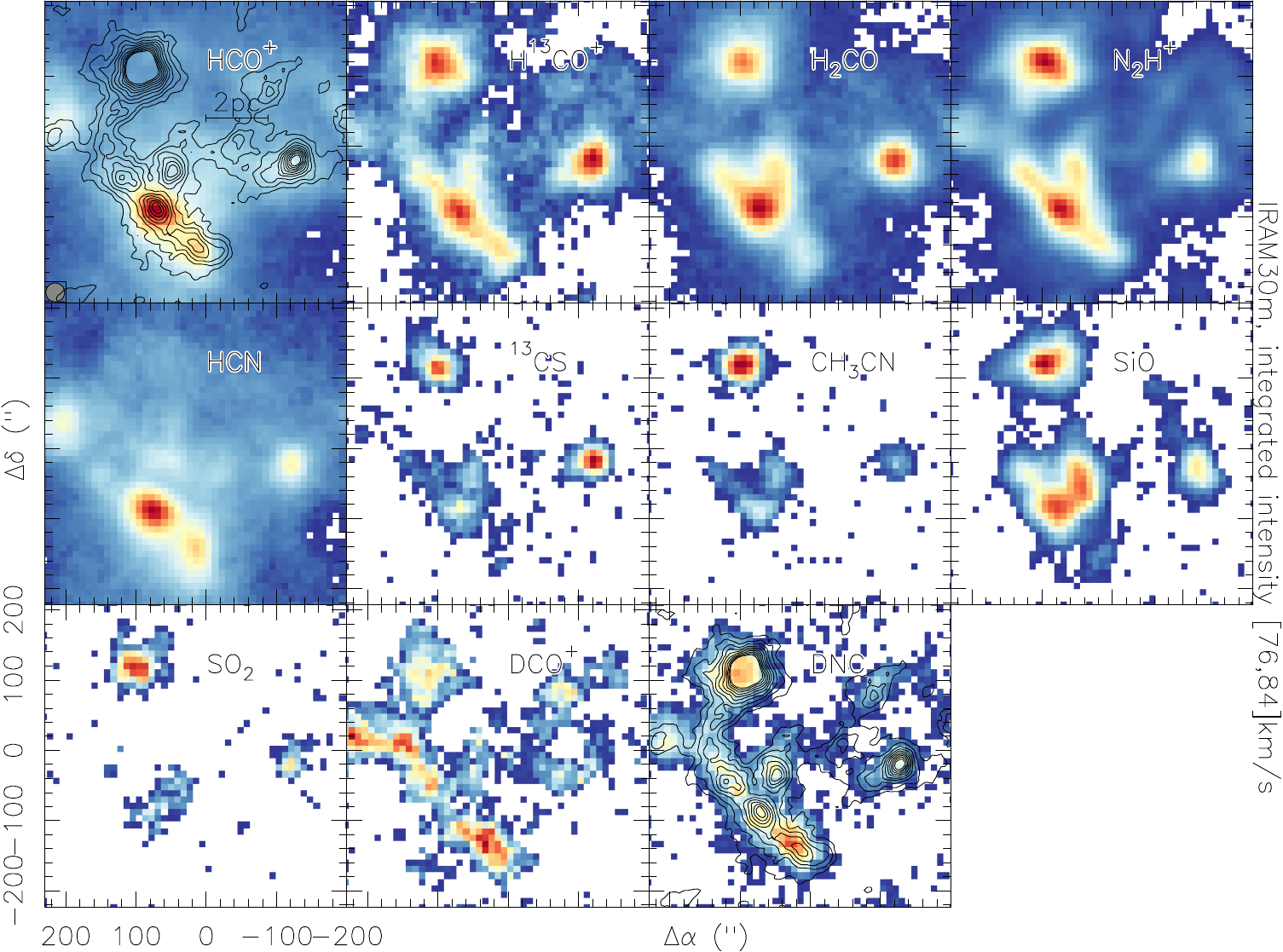}\\
\includegraphics[width=0.9\textwidth]{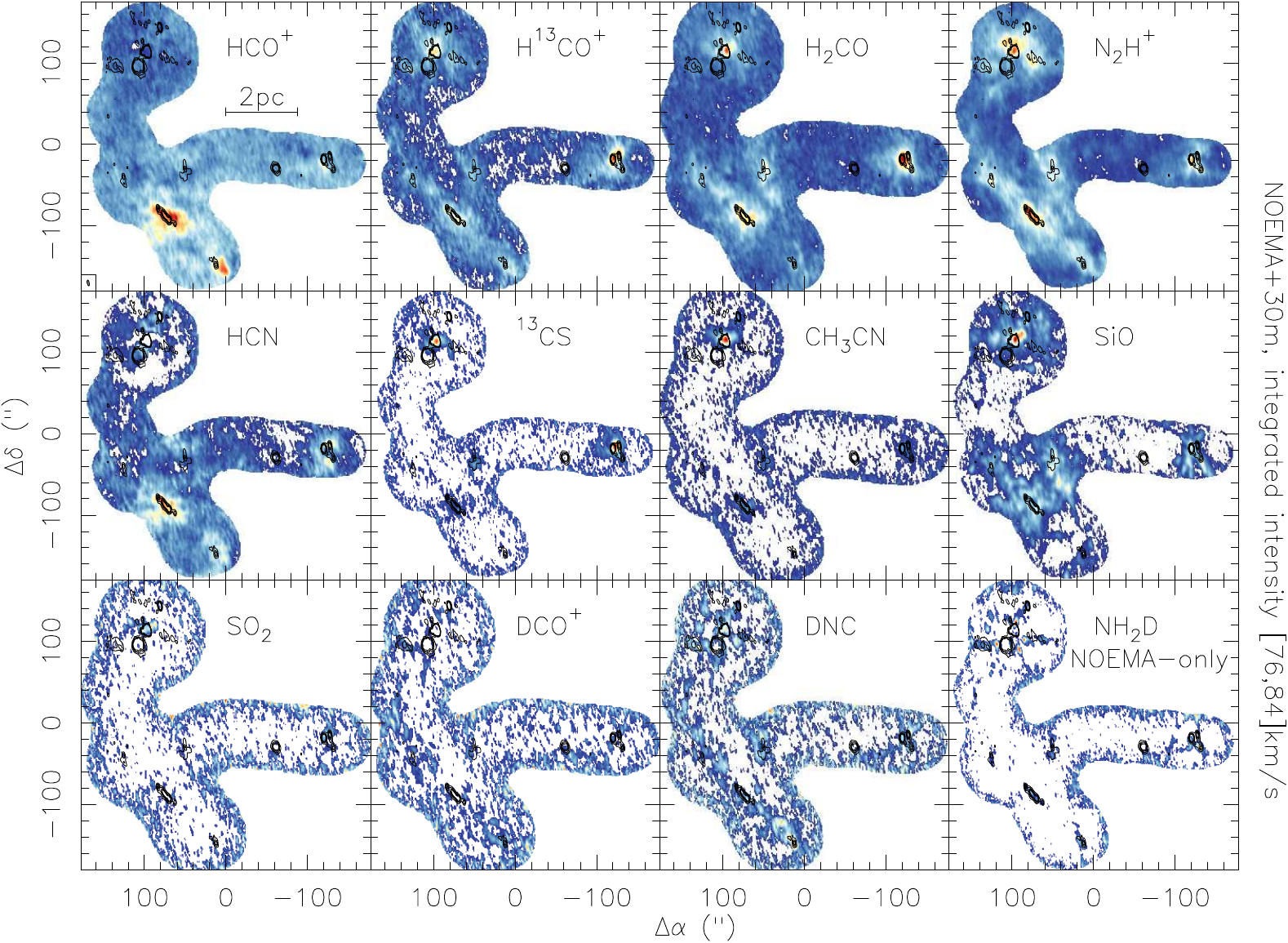}
\caption{Integrated intensity data for G28. Top:
  30\,m observations. Bottom: Merged
  NOEMA+30\,m data (except NH$_2$D, which shows NOEMA-only
  data as that line was not covered by the 30\,m observations). The
  integration range is 76 to 84\,km\,s$^{-1}$. All maps
  were created by clipping the data below the $3\sigma$ level;
  intensity levels were chosen for each panel separately to better
  highlight the emission features. The contours on the 30\,m data show
  870\,$\mu$m continuum \citep{schuller2009} in $3\sigma$ steps
  of 0.15\,Jy\,beam$^{-1}$. The contours on the NOEMA+30\,m data show the
  NOEMA-only 3.6\,mm continuum from 0.4 to to
  1.6\,mJy\,beam$^{-1}$ ($1\sigma \sim
  0.1$\,mJy\,beam$^{-1}$). Molecules are labeled in all panels, and
  the beam and scale bar are shown in the top-left panels.}
\label{mom0} 
\end{figure*} 

\begin{figure*}[htb]
\hspace{0.4cm} \includegraphics[width=0.87\textwidth]{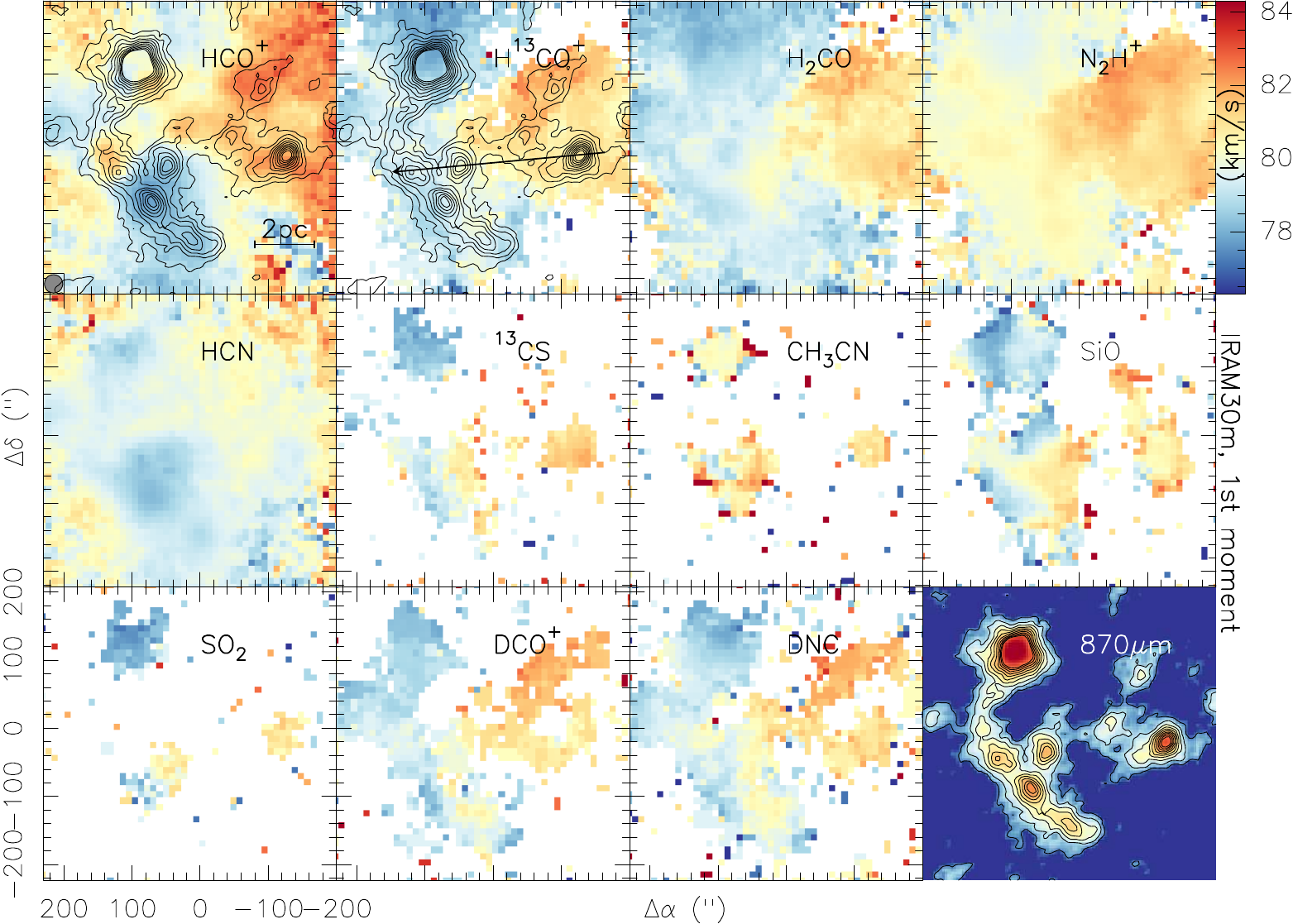}\\
\includegraphics[width=0.9\textwidth]{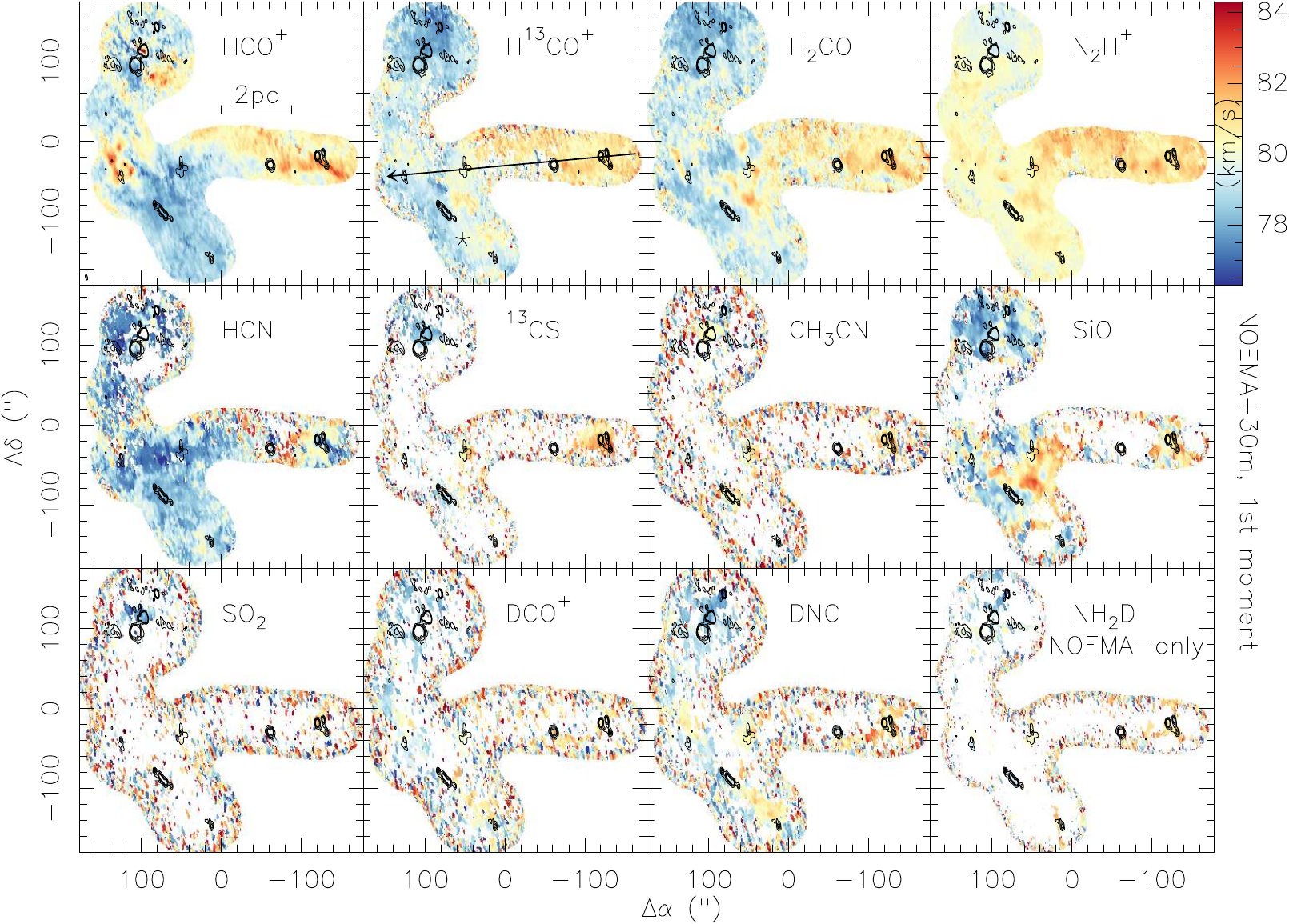}
\caption{First-moment maps (intensity-weighted peak velocities) for
  G28. Top: 30\,m observations. Bottom: Merged NOEMA+30m data (except NH
2
D, which shows NOEMA-only data as that line was
  not covered by the 30\,m observations). All maps were created by clipping the
  data below the $3\sigma$ level. The contours on the 30\,m data show
  870\,$\mu$m continuum \citep{schuller2009} in $3\sigma$ steps of
  0.15\,Jy\,beam$^{-1}$. The contours on the NOEMA+30\,m data show the
  NOEMA-only 3.6\,mm continuum from 0.4 to to
  1.6\,mJy\,beam$^{-1}$
  ($1\sigma\sim$0.1\,mJy\,beam$^{-1}$). Molecules are labeled in all
  panels, and the beam and scale bar are shown in the top-left
  panels. The H$^{13}$CO$^+$ panels outline the west-east pv cut
  presented in Fig.~\ref{pv}.}
\label{mom1} 
\end{figure*}  
\clearpage

The shock tracer SiO is an
interesting additional line that traces both compact and extended
emission but in a much wider velocity regime. We revisit this
in the following section.

\subsection{Kinematic properties}

For our kinematic analysis, Fig.~\ref{mom1} presents the first-moment
maps or intensity-weighted peak velocity for the same molecular lines
analyzed previously in the integrated emission maps. Again we show the
30\,m only data as well as the merged NOEMA+30\,m data (top and bottom
panels of Fig.~\ref{mom1}). All data consistently show a velocity
gradient from redshifted gas in the west to more blueshifted gas in
the east. Setting that velocity gradient in relation to previous
larger-scale work, this corresponds to the velocity gradient observed,
for example, in $^{13}$CO(3--2) from high to low Galactic latitudes in
Fig.~\ref{overview} \citep{beuther2020}.  This west-east velocity
gradient is visible in the extended 30\,m-only data as well as on the
small scales in the combined NOEMA+30\,m data. While the velocity
gradient is visible across the entire map in the single-dish data
(Fig.~\ref{mom1}), for the NOEMA high-resolution observations, we
focused only on the central west-east filament with the highest column
densities. The velocity ranges of the gradients appear slightly broader
in tracers of comparably lower density gas such as HCO$^+$ compared to
higher-density tracers such as N$_2$H$^+$ (Table \ref{lines}).

To get a better understanding of the velocity structures, we created
position-velocity (pv) cuts from west to east for the merged
NOEMA+30\,m data in the HCO$^+$ line, its $^{13}$C isotopolog, and
SiO. The exact outline of these pv cuts is shown in the H$^{13}$CO$^+$
panels of Fig.~\ref{mom1}. These high-resolution pv diagrams can be
directly compared to the larger-scale pv structure derived for the
[CI] and $^{13}$CO(3--2) emission in \citet{beuther2020}. Figure
\ref{pv} presents a compilation of these new and old data. While the
previous single-dish larger-scale observations cover an extent of more
that $10'$ across the main filament, our new NOEMA+30\,m
high-resolution data are more restricted and cover the gas flow from
the west toward the main filament and slightly beyond. Fig.~\ref{pv}
(top-right panel) outlines the different extents of the pv cuts. While
our new data do not cover the entire extent of the blueshifted gas
flow from the east (or low latitudes in the top-left panel of
Fig.~\ref{pv}), it nevertheless captures the convergent parts around
the main west-east filament (or high to low Galactic latitudes).

\begin{figure*}[h]
\includegraphics[width=0.99\textwidth]{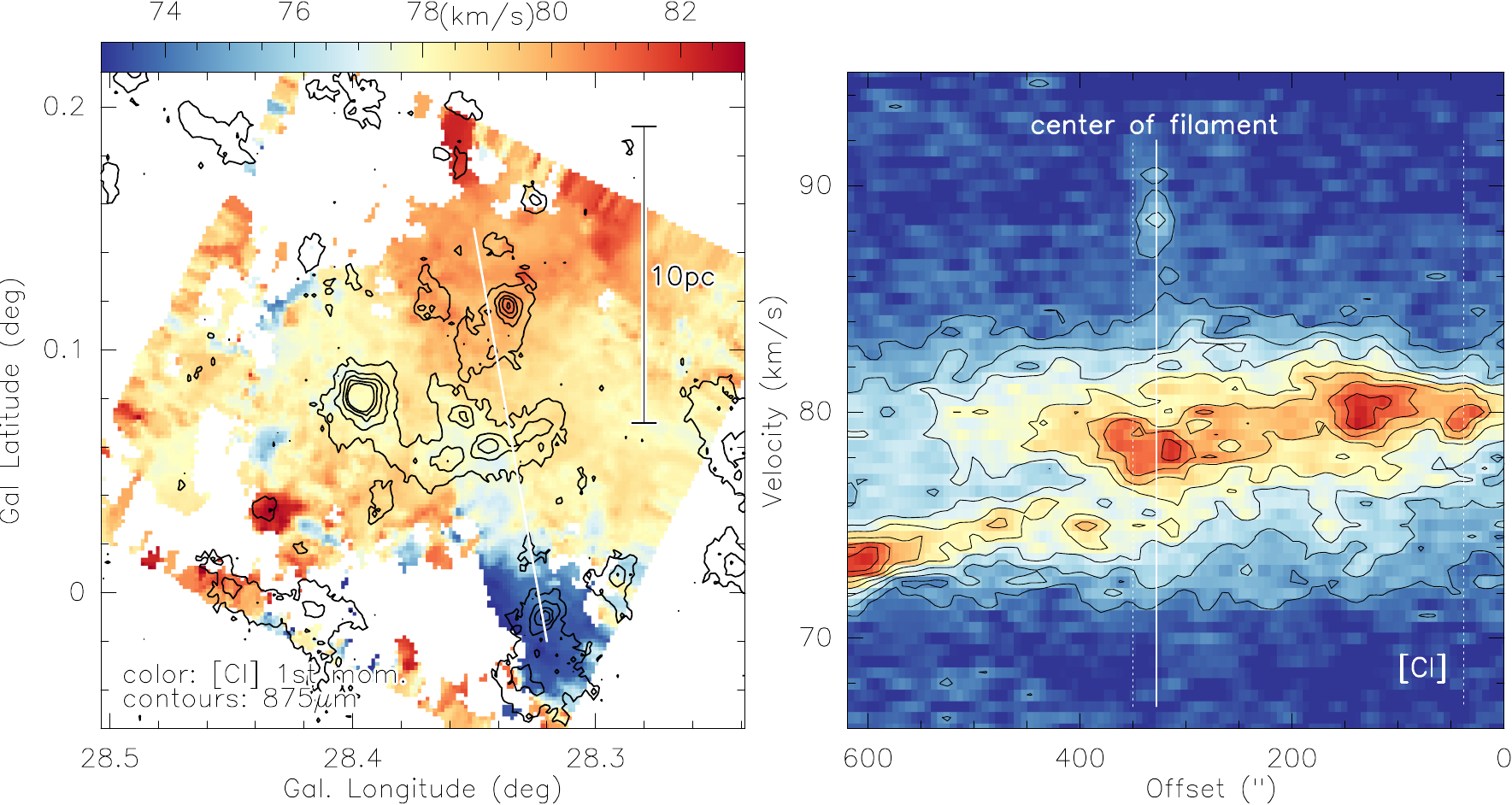}
\includegraphics[width=0.33\textwidth]{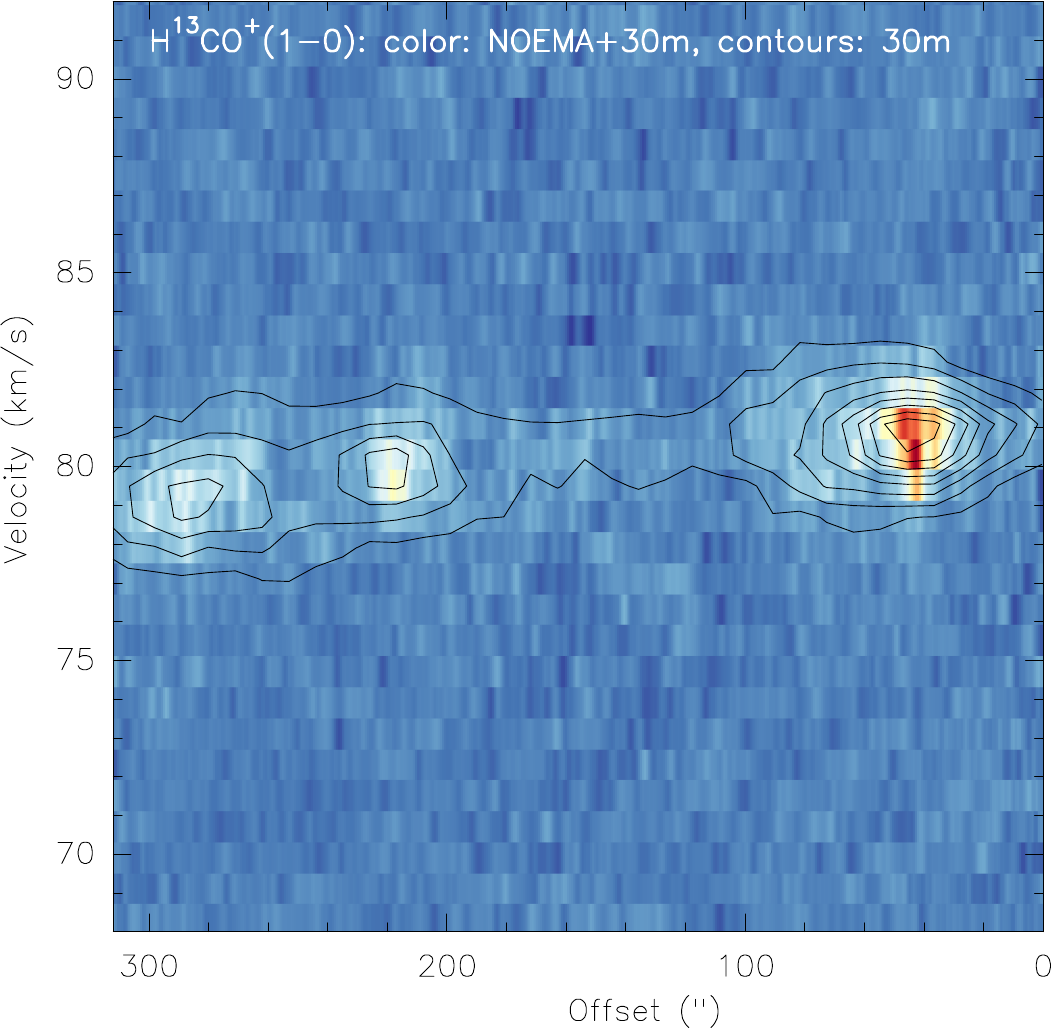}
\includegraphics[width=0.33\textwidth]{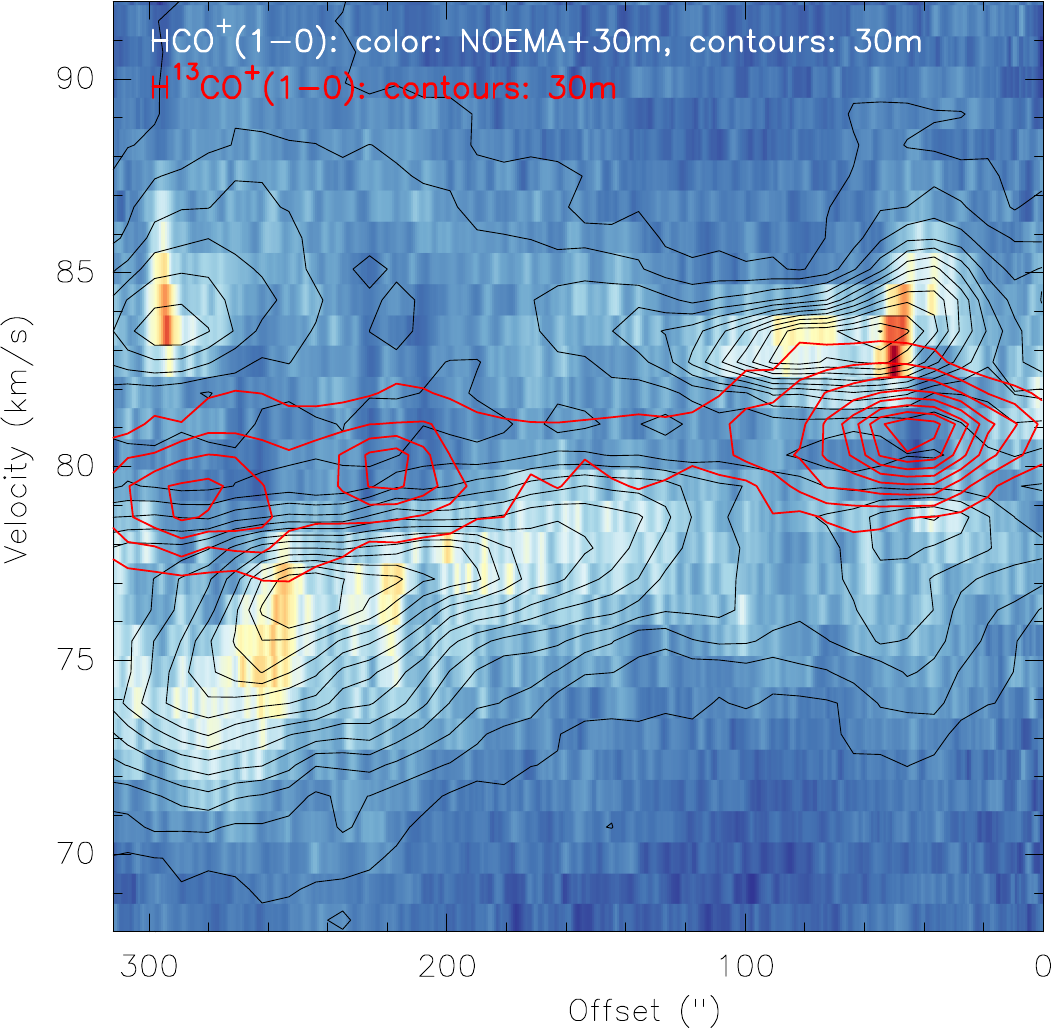}
\includegraphics[width=0.33\textwidth]{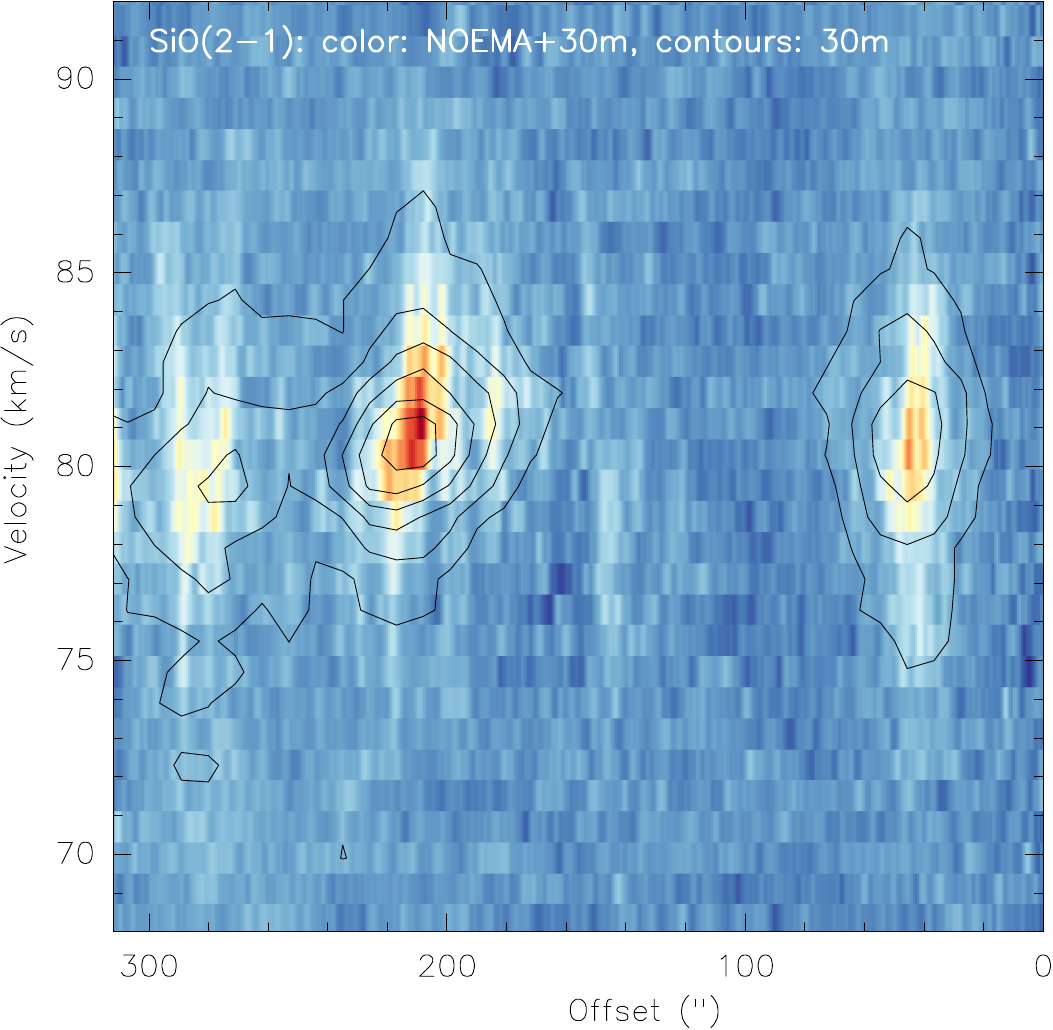}
\caption{Velocity comparison between large-scale APEX [CI] and
  small-scale NOEMA emission data. Top two panels: Reproduction of the [CI] first-moment map and pv cut
  from \citet{beuther2020}. Top-left: Color scale and contours
  showing the [CI] first-moment map and ATLASGAL 870\,$\mu$m emission
  (contour levels starting at $4\sigma$ and continuing in $8\sigma$ steps,
  $1\sigma$=50\,mJy\,beam$^{-1}$). A scale bar is shown as well. 
    Top-right: Position-velocity diagram along the white arrow
  in the top-left panel from high to low latitudes (high latitude
  offset 0). The full white lines mark the center position of the single-dish filament, and the dotted lines indicate the
  approximate extent of the corresponding pv cuts from the NOEMA
  data. Bottom row: Three pv diagrams in color, with contours along the west-east line (offset 0 indicates west) shown in
  Fig.~\ref{mom1} in H$^{13}$CO$^+$, HCO$^+$, and SiO. The color scale
  shows the corresponding NOEMA+30\,m data, and the contours show
  the 30\,m data only. The red contours in the middle HCO$^+$ panel
  also show the corresponding H$^{13}$CO$^+$ emission (30\,m-only
  data). Contour levels are in $4\sigma$ steps.}
\label{pv} 
\end{figure*} 

The new high-resolution, high-density data now reveal interesting new
kinematic information. While the lower-density tracer HCO$^+$
($n_{\rm eff}\sim 500$\,cm$^{-3}$) indeed shows blue-
to redshifted gas between $\sim$70 and $\sim$90\,km\,s$^{-1}$
(Fig.~\ref{pv}, bottom-middle panel), its higher-density tracing
$^{13}$C isotopolog H$^{13}$CO$^+$ ($n_{\rm eff}\sim 2.2\times
10^4$\,cm$^{-3}$) covers a much narrower velocity regime between
$\sim$77 and $\sim$93\,km\,s$^{-1}$ (Fig.~\ref{pv}, bottom-left
panel). Combining these two isotopologs into one panel
(Fig.~\ref{pv} bottom-middle panel), the $^{13}$C
isotopologs is stronger closer to the $v_{\rm lsr}$ where the main
isotopolog becomes optically thick. In addition to the main flow
signatures from red- to blueshifted gas from east to west, HCO$^+$
also traces comparably high-velocity gas toward all major sources and
in the region. These are signatures of the molecular outflows likewise visible in the SiO emission (Fig.~\ref{pv}, bottom-right panel).

Narrow linewidth SiO emission has in the past been suggested as a
potential tracer of low-velocity shocks from potential converging gas
flows (e.g., \citealt{jimenez-serra2010,cosentino2020}). As visualized
in the SiO pv cut (Fig.~\ref{pv} bottom-right panel), SiO does not show
significant narrow linewidth emission along the pv cut, but only
broad linewidth emission toward the three main sources along its
cut. Thess broad linewidths are likely caused by the many molecular
outflows known in the region (e.g., \citealt{kong2019}). The same
general SiO linewidth structure can also be seen in the second-moment
maps (intensity-weighted linewidths) presented in the appendix
Fig.~\ref{mom2}.

Although outlining the many outflows in the region is not the focus of
this paper, Fig.~\ref{sio} presents an overlay of the red- and
blueshifted SiO emission. One clearly finds high-velocity gas toward
G28P2/IRS2 and G28N in the north, G28P1/IRS1 in the center of the
filament, and G28P3 at the western end of the map. No high-velocity gas
is identified toward the more evolved H{\sc ii} regions IRAS18402-0403
and G28fil. Interestingly, our data also do not recover the outflow
known from the youngest regions G28S in the south of the filament
\citep{feng2016b,tan2016}. This non-detection in our data is most
likely caused by the region being the youngest and
weakest combined with our data's lower sensitivity and angular resolution compared to the previous detections mentioned.

\begin{figure}[htb]
  \includegraphics[width=0.49\textwidth]{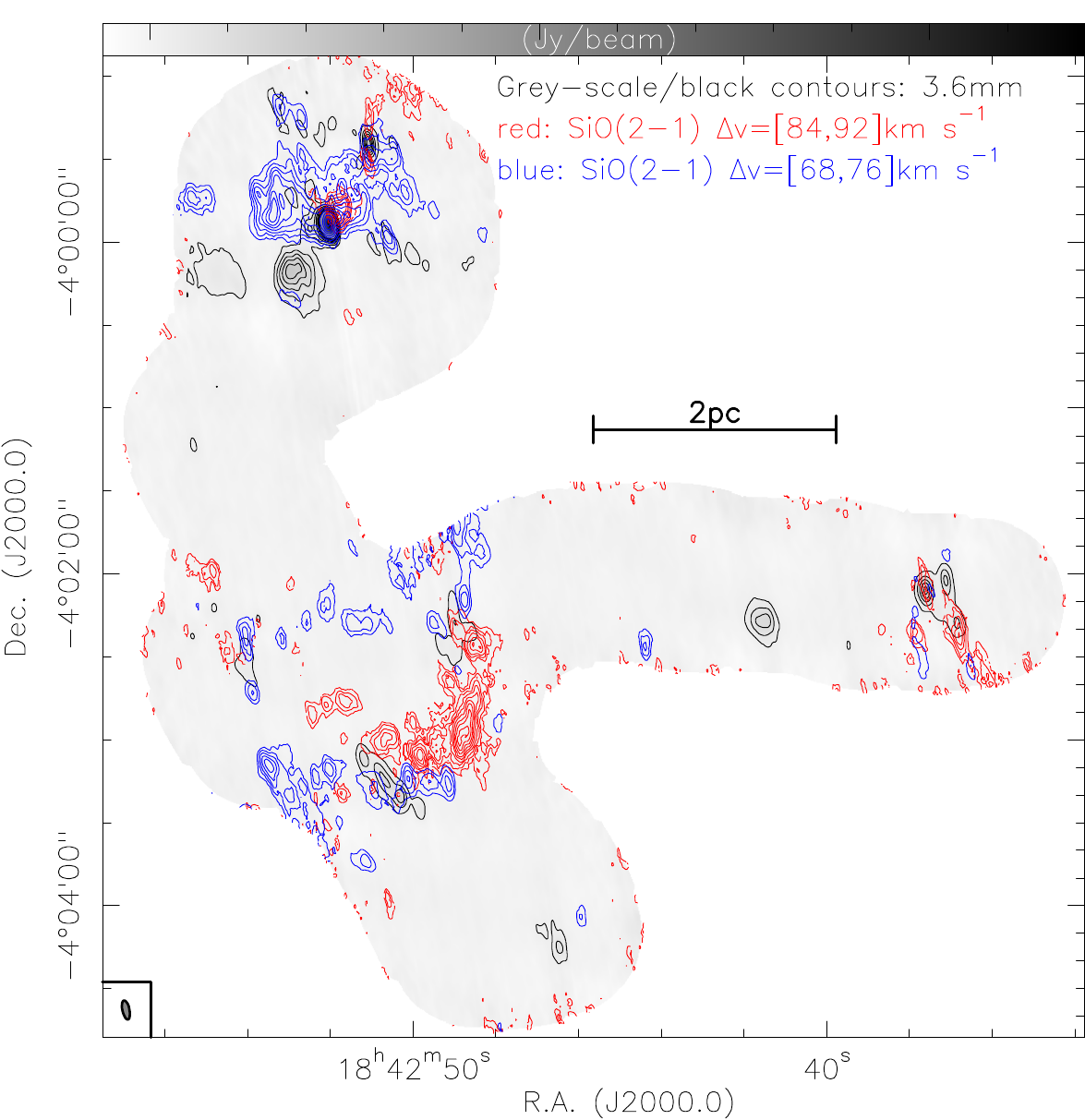}
\caption{High-velocity SiO(2--1) emission. The gray scale and contours
  show the 3.6\,mm continuum emission, as in Fig.~\ref{finding}. The red
  and blue contours show high-velocity SiO(2--1) emission with
  redshifted gas between 84 and 92\,km\,s$^{-1}$ and blueshifted gas
  between 68 and 76\,km\,s$^{-1}$, both in relative contour
  levels from 15 to 95\% of the integrated peak emission in
  10\% steps. A linear scale bar and the continuum synthesized beam are
  shown as well.}
\label{sio} 
\end{figure} 

\subsection{Gas flow parameters}

While other origins such as rotation could in principle also create
red- or blueshifted kinematic signatures, for G28, we interpret the
east-west velocity structure as arising from converging gas flows. The main
argument stems from the previous larger-scale analysis from
\citet{beuther2020} where the [CI] pv diagram is reproduced in
Fig.~\ref{pv} (top-right panel). This larger-scale pv diagram clearly
shows two separate velocity components -- blueshifted to the east and
redshifted to the west -- that only merge at the position of the
central north-south gas filament (longitudinal structure in Galactic
coordinates). Rotational signatures would produce a smooth velocity
gradient, whereas G28 clearly exhibits separate velocity structures
favoring the converging flow interpretation. Also, numerical
experiments typically indicate that rotation is not dominant on these
scales in turbulent flows (e.g., \citealt{smith2016}). Moreover, recent 3D dust studies of nearby regions found that motions in
the past attributed to rotation rather result from flow
convergence \citep{soler2013}. 

While the general west-east velocity gradient across the IRDC covers
the entire main north-south filament (Fig.~\ref{mom1}), in the
following we focus our analysis on the west-east filamentary structure, which is clearly visible in the (sub)millimeter continuum data. To now
quantify the mass flow rate $\dot{M}$ along the filamentary structure
from west to east, we follow the approach outlined, for instance, in
\citet{kirk2013,henshaw2014,wells2024,beuther2025b}:

\begin{eqnarray}
  \dot{M} = \frac{M_{\rm fil} \Delta v}{l_{\rm fil}} \times \frac{1}{{\rm tan}(i)} ~{\rm [M_{\odot}yr^{-1}]}.
  \label{mdot}
\end{eqnarray}  

Here, $M_{\rm fil}$ and $l_{\rm fil}$ are the mass and length of the
filament; $\Delta v$ is the velocity difference from the western end
of the filament to the eastern region. The final term $\frac{1}{{\rm
    tan}(i)}$ in Equation \ref{mdot} shows the dependence on the
inclination angle $i$ of the filamentary structure, which is unknown
in this case. \citet{wells2024} find a spread in the
flow rate distributions with inclination angle $i$ of roughly 1 order of magnitude
at full width half maximum (FWHM).

Since the NOEMA-only continuum data suffer from missing flux, we based
our estimate on the single-dish ATLASGAL 870\,$\mu$m dust continuum
and 30\,m-only spectral-line data. Using the same assumptions, as for
the total mass estimate from the ATLASGAL data in
Sect.~\ref{continuum_section}, we can estimate the mass in the entire
west-east filament (Fig.~\ref{continuum}) to roughly
4014\,M$_{\odot}$. With a length of the corresponding pv cut
(Fig.~\ref{mom1}) of $312''$, corresponding at a distance of 4.7\,kpc
to $\sim$7.1\,pc, a velocity difference in the H$^{13}$CO$^+$ data of
$\sim$2\,km\,s$^{-1}$ (from 81 to 79\,km\,s$^{-1}$, Fig.~\ref{pv}
bottom-left panel), and an assumed inclination angle of $i=45$\,deg,
the approximate mass flow rate along that filament is $\sim 1.15\times
10^{-3}$\,M$_{\odot}$yr$^{-1}$. While inclination uncertainties can
affect the flow rate (e.g., \citealt{wells2024}), we consider this
flow rate as an upper limit because it assumes all gas flows toward
the central filament. However, sources along the west-east
structure may already gather some of the material along the way.
Nevertheless, order-of-magnitude-wise, such flow rate values appear
plausible given that they feed a whole high-mass star-forming
region expected to form a massive cluster in the future. We
discuss that more in context in Sect.~\ref{flowrates}.

\begin{figure*}[htb]
  \includegraphics[width=0.99\textwidth]{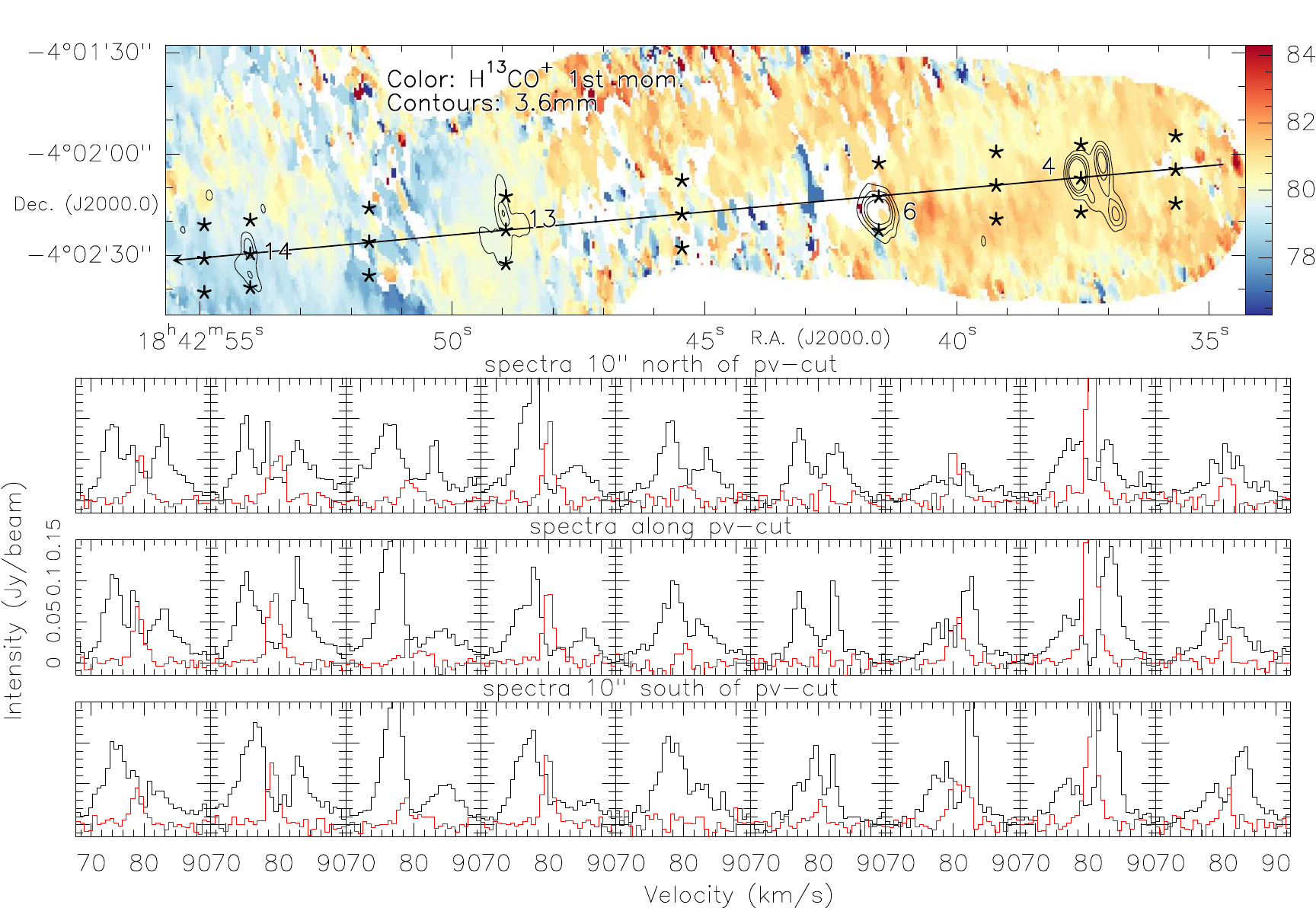}
\caption{Spectra along the west-east filament. Top: Finding chart with a color scale showing the
  H$^{13}$CO$^+$ first-moment map and contours showing the 3.6\,mm continuum in $4\sigma$ steps. The arrow indicates the pv cut from
  Fig.~\ref{pv}; the stars mark the positions where the
  spectra are extracted; and the main sources are labeled (see
  Fig.~\ref{finding}). Bottom: Corresponding HCO
+
 and H
13
CO
+
 spectra in black and red,
  respectively.}
\label{spectra} 
\end{figure*} 

A different way to investigate potential flow and infall motions is
via the classical asymmetric blue-skewed line profiles of optically
thick lines with respect to corresponding more Gaussian optically thin
counterparts (e.g.,
\citealt{myers1996,mardones1997,evans1999,devries2005,mottram2017,beuther2025}). While
the gas flow motions discussed so far largely correspond to motions
along the filamentary structures and in the plane of the sky
(depending on the inclination angle $i$, eqn.~\ref{mdot}), the motions
studied with the spectral profiles correspond to those along the
line of sight. Considering the general velocity gradient in the west-east
direction, we now investigate the gas motions along the line of sight
onto that structure. While we use the term ``flow'' for the west-east
gas stream, in the following, we use the term ``infall'' for the gas
motions along the line of sight.  The top panel of Fig.~\ref{spectra} shows a zoom-in on this west-east filament in the H$^{13}$CO$^+$ first-moment map and the 3.6\,mm continuum emission. We selected nine
positions along the pv cut, alternating between
positions offset from sources and positions toward sources. In
addition, we looked at spectra $10''$ north and south of the
main pv cut. All these 27 positions are marked by stars in Fig.~\ref{spectra}. The three lower spectral panels
in Fig.~\ref{spectra} present the corresponding HCO$^+$ and
H$^{13}$CO$^+$ spectra for these 27 positions (the top, middle, and bottom rows
correspond respectively to the north, central, and south positions marked in the top
panel). We selected the main isotopolog HCO$^+$ as the optically
thick tracer and its $^{13}$C isotopologs as its optically thin
counterpart.

In the picture of the asymmetric line profiles, blue-skewed profiles
of the optically thick line indicate infall motions (see Figs.~5 and 6
in \citet{evans1999} for an intuitive explanation). If one now
inspects the spectra along the pv cut in Fig.~\ref{spectra} (middle
row), one sees that these spectra are dominated by blue-skewed
structures. In particular, the spectra extracted toward the
intermediate positions offset from the sources largely exhibit the
blue-skewed signature. This is different than the on-source positions
toward sources 14, 6, and 4 (from east to west) where the profiles are
rather red-skewed and can be interpreted as outflow motions. This
makes sense for sources 14 and 4, which are known to drive
molecular outflows visible in the high-velocity SiO emission
(Fig.~\ref{sio}). Source 6 (or G28fil) is different since it already hosts a small H{\sc ii} region and is hence likely to have
terminated any infall or accretion process. Therefore, we do
not expect to see any infall motions toward the sources driving outflow and
hosting the H{\sc ii} region. An interesting case is also exhibited
by the spectra toward source 13, which indeed show the blue-skewed infall
signature while simultaneously driving an outflow as visualized in the
SiO emission (Fig.~\ref{sio}). In the infall picture, the outflow can
lie nearly in the plane of sky without affecting the line-of-sight
signatures too much. Regarding the off-source positions, only that
between sources 6 and 4 shows clear red-skewed profiles.

If one leaves the central cut but moves $10''$ to the north and south (Fig.~\ref{spectra}, top and bottom spectral rows), one also
identifies a clear dominance of the blue-skewed infall-indicating
profiles. Altogether, these spectral structures strongly
indicate infall motions toward the west-east filamentary
structure.

To quantify these infall motions along the line of sight, we can
follow the classical two-layer model described originally in
\citet{myers1996}. This model assumes two gas layers of equal
temperature and velocity dispersion $\sigma$ that collapse toward a
central region. The front and back layers correspond to the red- and
blueshifted gas. Following their equation 9, one can estimate the
infall velocity $v_{\rm in}$ as
\begin{eqnarray}
v_{\rm in} \approx \frac{\sigma^2}{v_{\rm red}-v_{\rm blue}}{\rm ln}\left(\frac{1+eT_{\rm BD}/T_{\rm D}}{1+eT_{\rm RD}/T_{\rm D}}\right),
\end{eqnarray}
where $v_{\rm red}$ and $v_{\rm blue}$ correspond to the red- and
blueshifted peak positions, $T_{\rm D}$ is the intensity at the
position of the dip, and $T_{\rm RD}$ and $T_{\rm BD}$ are the
intensity differences between the red- and blueshifted peaks and the
dip in the spectra. Using the spectral parameters measured in
H$^{13}$CO$^+$ and HCO$^+$ toward source 13 (fourth spectrum from the left in
the middle row of Fig.~\ref{spectra}, Table \ref{infall}), we
infer an infall velocity $v_{\rm in}$ of
$\approx$0.11\,km\,s$^{-1}$. For comparison, the thermal sound speed
of the molecular gas at 20\,K of $\approx$0.27\,km\,s$^{-1}$
corresponds to a thermal FWHM linewidth of H$^{13}$CO$^+$ at 20\,K of
$\approx$0.17\,km\,s$^{-1}$. Hence, the inferred infall velocity along
the line of sight toward this source is subsonic.

\begin{table}[htb]
  \caption{Infall parameters}
  \label{infall}
  \begin{tabular}{lrrrrrrrr}
    \hline \hline
    \# & $\Delta v$ & $v_{\rm blue}$ & $v_{\rm red}$ & $T_{\rm B}$ & $T_{\rm R}$ & $T_{\rm D}$ & $v_{\rm in}$ & $\dot{M}_{\rm in}$ \\
       & $\frac{\rm km}{\rm s}$  & $\frac{\rm km}{\rm s}$  & $\frac{\rm km}{\rm s}$ & $\frac{\rm Jy}{\rm beam}$ & $\frac{\rm Jy}{\rm beam}$ & $\frac{\rm Jy}{\rm beam}$ & $\frac{\rm km}{\rm s}$  & $\frac{10^{-4}{\rm M_{\odot}}}{\rm yr}$ \\
    \hline
    13 & 2.2 & 77.0 & 86.7 & 0.117 & 0.042 & 0.018 & 0.11 & 0.45 \\
    8  & 2.6 & 77.9 & 85.1 & 0.622 & 0.062 & 0.003 & 0.38 & 2.7 \\
    11 & 3.0 & 78.7 & 82.7 & 0.499 & 0.052 & 0.023 & 1.01 & 4.9 \\
    \hline \hline
  \end{tabular}
  Notes: The value $\Delta v$ represents the FWHM from H$^{13}$CO$^{+}$. All other
  spectral parameters are from HCO$^+$. The values $v_{\rm blue}$, $v_{\rm
    red}$. $T_{\rm B}$, and $T_{\rm R}$ represent the blue- and redshifted
  peak-velocities and peak intensities, and $T_{\rm D}$ represents the
  intensity at the dip-velocity.
\end{table}

With the estimated infall velocity along the line of sight and the
previously derived column density (at 20\,K, Table \ref{cont}), we can
also estimate an infall rate $\dot{M}_{\rm in}$ along the
line of sight with
\begin{eqnarray}
  \dot{M}_{\rm in} = N \times v_{\rm in} \times r,
\end{eqnarray}
where $r$ corresponds to the radius which we assume as the mean FWHM
of the $\sim$4.8$''$ beam. With the given values (Tables \ref{cont}
\& \ref{infall}), we infer an approximate infall rate
along the line of sight $\dot{M}_{\rm in}$ of $\approx 4.5\times
10^{-5}$\,M$_{\odot}$yr$^{-1}$ for source 13. This infall rate along the
line of sight is roughly a factor of 25 lower than the flow rate along
the west-east filamentary structure estimated above. This large
difference provides additional evidence that the gas flow
is indeed dominated by a west-east converging gas flow, with much
smaller flow contributions along the line of sight.

In addition to the flow and infall velocities associated with the
west-east filamentary structure, we can also estimate the infall
velocities $v_{\rm in}$ and infall rates $\dot{M}_{\rm in}$ for a few
other regions -- in particular, the two prominent star-forming regions
G28P1 (source 8) and G28S (source 11) along the prominent north-south
filament. Figure \ref{spectra_others} presents the corresponding
HCO$^+$ and H$^{13}$CO$^+$ spectra, which show the blue-skewed spectra indicative of infall. Deriving the spectral
parameters (Table \ref{infall}), the estimated infall velocities and
infall rates are all higher than those measured toward source 11
along the west-east filament. More precisely, we find $v_{\rm
  in}$ and $\dot{M}_{\rm in}$ toward G28P1 and G28S of
0.38/1.01\,km\,s$^{-1}$ and $2.7\times 10^{-4}$/$4.9\times
10^{-4}$\,M$_{\odot}$yr$^{-1}$, respectively. This implies that the
infall velocities toward these sources are rather
supersonic. Furthermore, infall rates along the line of sight in the
prominent north-south filament appear roughly 1 order of magnitude
greater than those along the converging west-east flow filament. We
discuss this in more detail in Sect.~\ref{flowrates}.

\section{Discussion}
\label{discussion}

\subsection{Flow rates}
\label{flowrates}

Aiming to better understand the converging gas flow leading to
the prominent G28 IRDC, we can now quantitatively characterize the
different flow contributions. While earlier studies of the more
diffuse gas on larger spatial scales (up to 15\,pc) estimated
large-scale flow rates on the order of a few times
10$^{-5}$\,M$_{\odot}$yr$^{-1}$ \citep{beuther2020}, the flow rates we
now infer along the west-east filament ($\sim$7.1\,pc length) in the
dense-gas-tracing line H$^{13}$CO$^+$(1--0) are higher around
$1.14\times 10^{-3}$\,M$_{\odot}$yr$^{-1}$. While these flow rates
should be considered an upper limit -- since sources along the
filament can be fed along the flow -- the higher inferred flow rate
of the dense gas closer to the central infrared-dark filament should
be real. While some models predict even opposite behaviors -- i.e.,
decreasing flow-rates closer to the center because the large-scale
flow can split into several small-scale flows (e.g.,
\citealt{padoan2020}) -- acceleration of gas by increased gravitational
pull may also increase the flow rates, as observed here. In general,
such high flow rates are reasonable since the converging gas flow
appears to feed an entire infrared-dark filament and high-mass
clusters typically gain most of their mass within several
hundred thousands of years (e.g.,
\citealt{yorke2002,mckee2003,kuiper2011,bonnell2011,rosen2022,oliva2023b}).

The mass flow rates around $10^{-3}$\,M$_{\odot}$yr$^{-1}$ correspond
to the standard deviation of mass flow rates obtained recently for the
local ISM within 1.25\,kpc from the Sun (Section
5.4 in \citealt{soler2025}).  If one assumed that the environment of
G28 were similar to the local ISM (unlikely, since G28 lies in
the Galactic midplane forming high-mass stars), one might conclude
that G28 formed from a typical fluctuation in the interstellar flows
encountered in the ISM. However, since the signatures of
the large-scale gas flow presented here and in \citet{beuther2020}
clearly point to converging gas flows, we instead infer that G28 is
undergoing a converging mass flow comparable to those
found in the local ISM.

Focusing on the west-east filament, the flow is dominated by
converging gas along that structure with a much smaller flow-rate
contribution along the line of sight. This reconfirms the picture of
a converging gas flow with a west-east orientation.

This differs from estimated flow rates along the line of sight in
the prominent north-south filament, which should result from the
west-east converging gas flow (see also
\citealt{beuther2020}). Toward sources at the heart of the infrared-dark cloud, in particular G28P1 or G28S, we infer mass infall rates on
the order of a few times $10^{-4}$\,M$_{\odot}$yr$^{-1}$. Such infall
rates toward individual sources are again reasonable for the formation
timescales of high-mass stars of a few hundred thousand years.

It is interesting to note that the flow rates along the line of sight
differ by roughly an order of magnitude along the west-east filament
compared to the values estimated toward the sources in the prominent
north-south infrared-dark filament. This difference may indicate that
the gas flow in the west-east direction along the converging gas flow is
indeed dominated by this directional gas flow, whereas in the final
formed structure -- the north-south filament -- the flow directions become more multidirectional. This proposed difference between
flow rates in converging gas flows and those toward the final
star-forming clumps needs further investigation by future studies.

\subsection{Decoupling of gas and dust temperatures}
\label{decouple}

While the gas and dust temperatures presented in
Sect.~\ref{continuum_section} and Fig.~\ref{temp} are already required
for accurate mass and column density estimates, the inferred differences
between dust- and CH$_3$CN-derived gas temperatures is also an
important topic in themselves. At the given high critical and effective
densities of CH$_3$CN ($\sim 10^5$ and $\sim 4-7\times
10^4$\,cm$^{-3}$ at 20\,K, respectively;
\citealt{schoeier2005,shirley2015}, Table \ref{lines}), one would
expect the gas and dust to be well coupled and hence at roughly the same
temperature (e.g., \citealt{stahler2005}). We note that our gas
temperature estimates are in qualitative agreement with the ones
presented in \citet{wang2008}, who used NH$_3$ to derive cold
temperatures ($\sim$16\,K) near G28PI/IRS1 and elevated temperatures
($\sim$27\,K) around G28P2/IRS2. As shown in Table \ref{cont} and
Fig.~\ref{temp}, wherever CH$_3$CN is detected, the inferred gas
temperatures are systematically higher than the inferred dust temperatures. While
the temperature differences decrease when the CH$_3$CN
data are smoothed to the same spatial resolution ($12''$) as the dust temperature
map, systematic differences nevertheless remain
(Fig.~\ref{temp}). For all sources except source 7, the
CH$_3$CN-derived temperatures remain either higher or nearly the same,
with the largest difference being almost a factor of 2 for source 1 (Table
\ref{cont}). These differences may arise from technical factors as well as those of physical origin.

Regarding technical factors, the dust temperatures are based on
Herschel far-infrared data at wavelengths longer than
70\,$\mu$m. Assuming blackbody emission, such long wavelength data are
by definition only sensitive to cold material, typically between 10
and 40 K. Temperatures barely above 50\,K can be estimated with
such kind of data (e.g., \citealt{marsh2017}). In contrast,
CH$_3$CN contains whole $k$-ladders with lines at various excitation
temperatures. In the case of CH$_3$CN$(4_k-3_k)$ and
CH$_3$CN$(5_k-4_k)$, the $k-$lines with $k=0...4$ cover excitation
energies from below 20 to above
100\,K\footnote{\href{https://splatalogue.online}{SPLATALOGUE
  webpage}}. Therefore, CH$_3$CN is sensitive to a much broader range of
temperatures. Contrary to this, since dust emission occurs at all
column and volume densities, the dust temperature map in
Fig.~\ref{temp} covers the entire mapped area, whereas the CH$_3$CN is
only detected toward dense sources due to its high critical
density. Hence, while the dust temperature map is more useful for obtaining
an overall picture of the entire region, the CH$_3$CN temperature maps
may better represent the temperatures of the dense sources. In
addition, since the dust continuum emission is sensitive to
the density and temperature structure along the entire line of sight --
including foreground and background -- it may not reliably reflect local
gas conditions (see also \citealt{feng2019}).

Decoupling of gas and dust temperatures at high densities has in the
past been mainly observed and discussed in the central molecular zone
(CMZ) of the Milky Way (e.g.,
\citealt{marsh2016,immer2016,ginsburg2016,krieger2017,tang2021,henshaw2023}).
The temperature differences there are slightly higher, with dust
temperatures between 20--50\,K and a corresponding gas temperature
range of 50--100\,K \citep{henshaw2023}. In comparison, for source
1 we find a difference between gas and dust temperatures around $\sim
2$. For most other sources, however, the difference is typically $\leq 10$\,K
(Table \ref{cont}). An additional difference between the CMZ and the
G28 cloud is that toward the CMZ, the gas and dust temperature
differences have been observed with single-dish telescopes over
comparably large spatial scales (e.g.,
\citealt{ginsburg2016,immer2016}), whereas in the G28 IRDC, they are
rather localized around protostellar regions.

Nevertheless, a decoupling of gas and dust temperatures at high
densities also appears in this prototypical IRDC,
G28. Following the discussions of the temperature decoupling in the
CMZ (see summary in \citealt{henshaw2023}), radiative heating is
inefficient due to poor coupling between radiation and gas (e.g.,
\citealt{clark2013,ao2013,ginsburg2016}). The main candidates for
explaining the different gas and dust temperatures are mechanical
heating, i.e., transformation from kinetic to thermal energy at the
end of the turbulent cascade via shocks or cosmic ray heating (e.g.,
\citealt{pan2009,papadopoulos2010}). Studies indicate that mechanical
heating appears the more likely candidate for the CMZ (e.g.,
\citealt{ginsburg2016,immer2016}). Temperature differences between gas
and dust on core scales averaging around a factor of 1.7 have recently
also been reported from the ALMA-IMF project \citep{motte2025}.

Since the G28 IRDC formed through a converging gas flow, shocks and the
conversion of kinematic flow energy into thermal energy also appear a
very plausible gas heating process for this region. Nevertheless,
since we do not know the cosmic ray ionization rate in G28, we cannot
exclude contributions from cosmic rays to the higher gas
temperatures. Most likely, a combination of both effects may explain
the observed temperature differences. Future observations of the
cosmic ray ionization rate in this and other regions, as well as
simulations of converging flows and their associated energy
dissipation and conversion processes, will help shed more light on
the different processes in such typical Milky Way clouds.

\section{Conclusions and summary}
\label{summary}

Observing the prototypical infrared-dark cloud G28 at a comparably high
angular resolution (approximate beam size of $4.8''$ or $\sim$22500\,au)
over large spatial scales ($\sim$81\,pc$^2$) in a series of dense
gas tracers and the 3.6\,mm continuum emission, we can dissect the
kinematic, fragmentation, and physical properties of the region in
great depth. While presenting overall data from NOEMA and the IRAM 30\,m
observatory, we focused on the kinematic analysis
of the converging gas flow and a partial decoupling of the gas and
dust temperatures.

Our analysis confirms the converging west-east gas flows in many dense
gas tracers, and we can quantitatively estimate a overall gas flow
rate around $\sim 10^{-3}$\,M$_{\odot}$yr$^{-1}$. While this may represent an
upper limit since some condensations along the flow potentially use
up some of the gas, the flow rate along the converging gas flow is
roughly a factor of 25 greater than infall rates measured in that region
along the line of sight. This difference confirms the dominance of the
west-east converging gas motions over gas motions along the
line of sight. Interestingly, infall rates along the line of sight
within the main north-south filaments are on the order of a few times
$10^{-4}$\,M$_{\odot}$yr$^{-1}$, significantly higher than the
line-of-sight infall rates estimated along the west-east gas
flow. This indicates that, where the flows converge and most
star formation takes place, the initially more directed
converging gas flows may convert into more multidirectional infall
motions.

A comparison of dust temperatures measured from Herschel
far-infrared data with gas temperature estimates from our new
CH$_3$CN data reveals differences between the two. The gas
temperatures are typically higher than the dust
temperatures. This is surprising because the high critical densities
of CH$_3$CN ($\sim$10$^5$\,cm$^{-3}$) lead to the expectation
thermal coupling between gas and dust. These differences are partly
reminiscent of the results otherwise mainly found in the CMZ of the
Milky Way. Following the discussions on different gas and dust
temperatures in the CMZ, mechanical heating and/or cosmic ray heating
may explain the differences observed in the G28 IRDC as well. Since
this IRDC results from a converging gas flow, the conversion of
kinematic flow energy into thermal gas energy may indeed explain the
higher gas temperatures compared to the dust temperature counterparts.

\begin{acknowledgements} 
  We thank the referee for the insightful comments improving the
  paper. This work is based on observations carried out
  with the IRAM NOEMA Interferometer and the IRAM 30\,m
  telescope. IRAM is supported by INSU/CNRS (France), MPG (Germany)
  and IGN (Spain). D.~S. has received funding from the European
  Research Council (ERC) under the European Union’s Horizon 2020
  research and innovation programme (PROTOPLANETS, grant agreement
  No. 101002188).
\end{acknowledgements}

\newpage

\begin{appendix}

\section{Additional figures}
  
\begin{figure*}[htb]
\hspace{0.4cm} \includegraphics[width=0.87\textwidth]{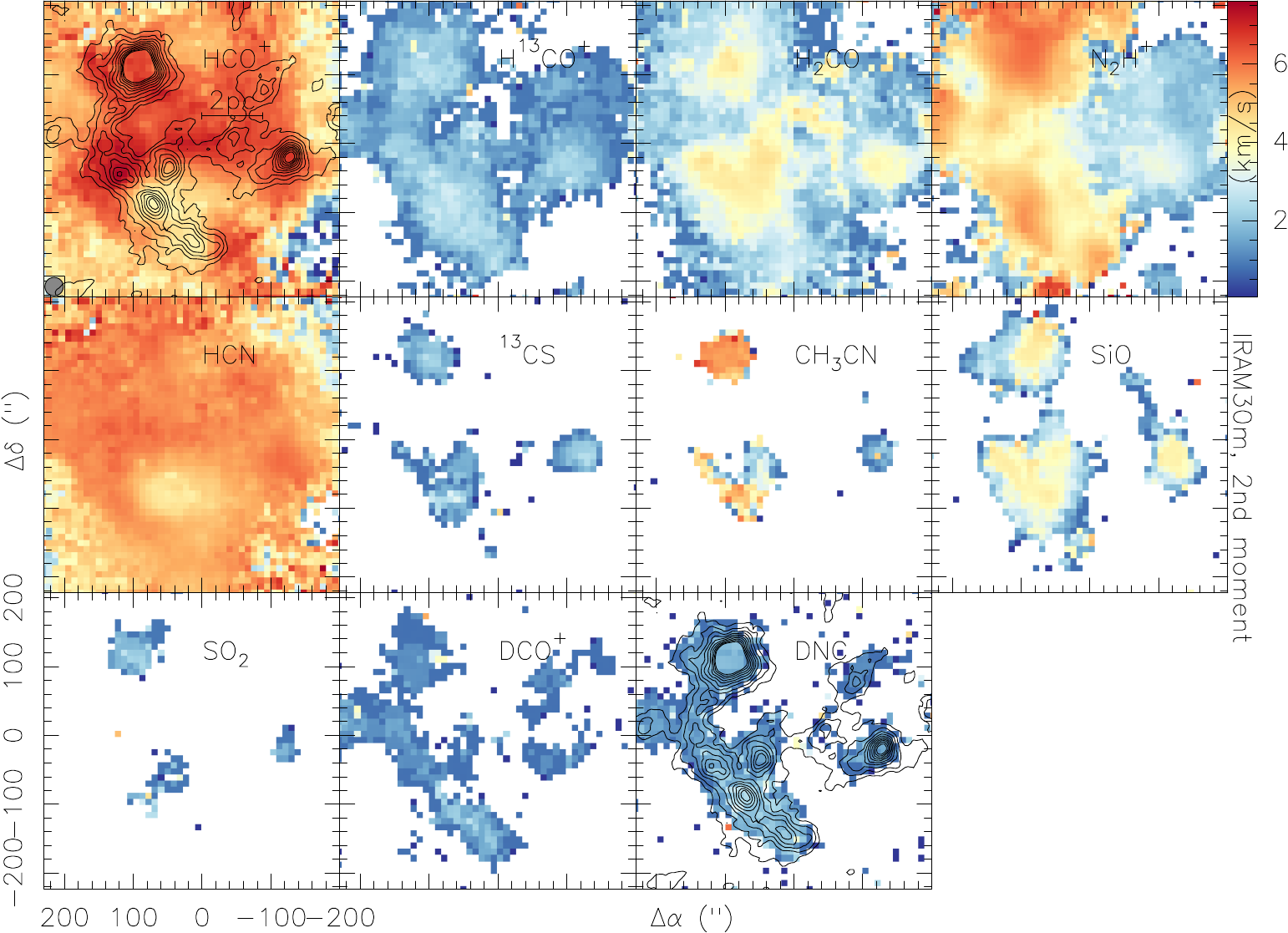}\\
\includegraphics[width=0.9\textwidth]{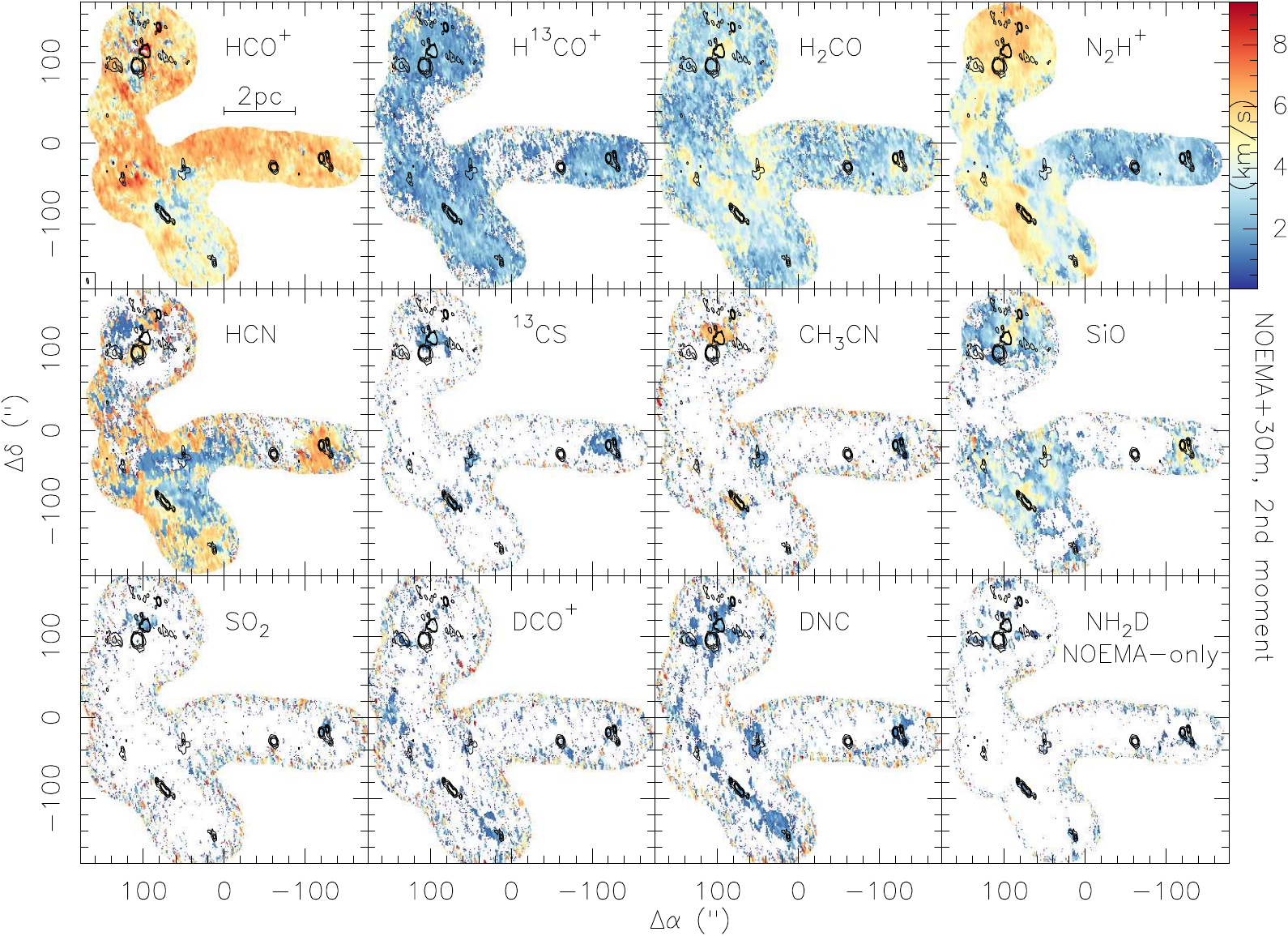}
\caption{Second moment maps (intensity-weighted velocity dispersion)
  for G28. Top: 30\,m observations.
  Bottom: Merged NOEMA+30\,m data (except NH
2
D showing NOEMA-only data as that line
  was not covered with the 30\,m observations). All maps were created by
  clipping the data below the $3\sigma$ level. The contours on the
  30\,m data show 870\,$\mu$m continuum \citep{schuller2009} in
  $3\sigma$ steps of 0.15\,Jy\,beam$^{-1}$. The contours on the
  NOEMA+30\,m data show the NOEMA-only 3\,mm continuum from 0.4 to 1.6\,mJy\,beam$^{-1}$ ($1\sigma \sim
  0.1$\,mJy\,beam$^{-1}$). Molecules are labeled in all panels, and
  the beam and scale bar are shown in the top-left panels.}
\label{mom2} 
\end{figure*}

\begin{figure}[htb]
\includegraphics[width=0.49\textwidth]{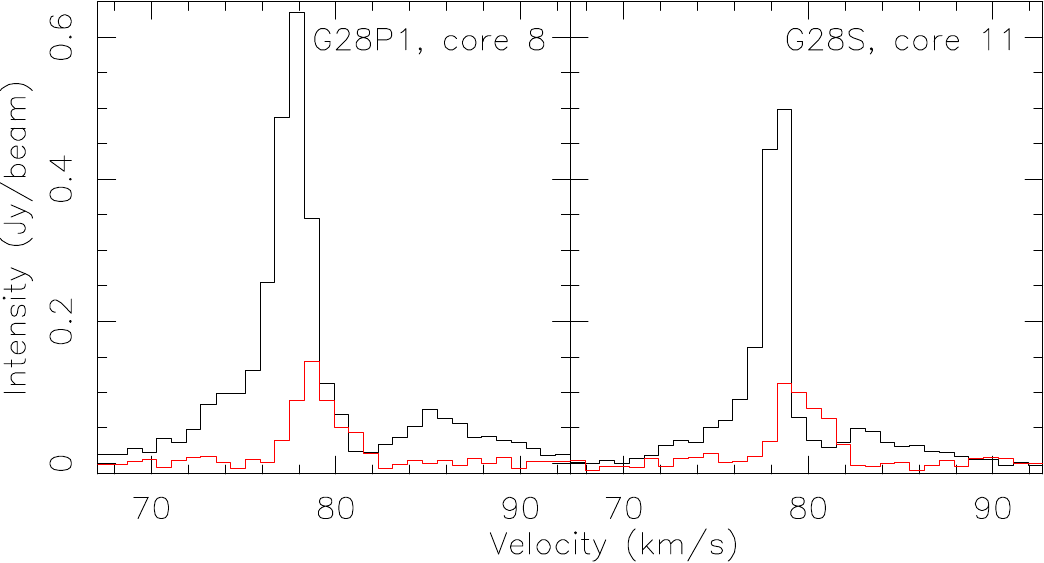}
\caption{Spectra toward G28P1 (source 8) and G28S (source 11). Black and
  red correspond to the HCO$^+$ and H$^{13}$CO$^+$ spectra,
  respectively.}
\label{spectra_others} 
\end{figure} 

\end{appendix}

\end{document}